\documentclass[12pt]{article}  
\usepackage[a4paper,textwidth=490.0pt,textheight=703.1pt]{geometry}
\usepackage{slashed}
\usepackage{graphicx}
\usepackage{epsfig}
\usepackage{amsmath}
\usepackage{amssymb}
\usepackage{ulem}
\usepackage{mdwlist}
\usepackage{color}                                     
\usepackage{float}
\usepackage[rflt]{floatflt}
\usepackage{slashed}
\usepackage{cite}     

\usepackage{array}

\usepackage[english]{babel}
\usepackage[latin1]{inputenc}
\usepackage[T1]{fontenc}
\usepackage{ae}

\usepackage{url}

\usepackage{amsmath, amsthm, amssymb}

\catcode`,\active

\catcode`\,12

\newtheorem{thm}{Theorem}[section]

\usepackage{array}

\usepackage[english]{babel}
\usepackage[latin1]{inputenc}
\usepackage[T1]{fontenc}
\usepackage{ae}

\newtheorem{definition}[thm]{Definition}

\newcommand{\gsim}{\raisebox{-0.07cm   }
{$\, \stackrel{>}{{\scriptstyle\sim}}\, $}}
\newcommand{\GeV}{\rm GeV}

\newcommand{\Li}{{\rm Li}}

\usepackage{rotating}

\usepackage{graphicx}

\newcounter{mmacnt}
\def\restartmma{\setcounter{mmacnt}{0}}
\restartmma \catcode`|=\active
\def|#1|{\mathrm{#1}}
\catcode`|=12
\newenvironment{mma}{
 \par\smallskip
 \catcode`|=\active
 \parskip=0pt\parindent=0pt 
 \small
 \def\In##1\\{%
   \def\linebreak{\hfill\break\null\qquad}%
   \refstepcounter{mmacnt}
   \hangindent=2.5em\hangafter=0
   \leavevmode
   \llap{\tiny\sffamily In[\arabic{mmacnt}]:=\kern.5em}%
   \mathversion{bold}\footnotesize$\displaystyle##1$\normalsize
   \mathversion{normal}\par
 }%
 \def\Print##1\\{%
   \def\linebreak{\hfill\break}%
   \hangindent=2.5em\hangafter=0
   \leavevmode ##1\par}%
 \def\Out##1\\{%
   \def\linebreak{$\hfill\break\null\hfill$}%
   \kern\abovedisplayskip\par
   \hangindent=2.5em\hangafter=0
   \leavevmode
   \llap{\tiny\sffamily Out[\arabic{mmacnt}]=\kern.5em}
   \footnotesize$\displaystyle##1$\normalsize\hfill\null\par
   \kern\belowdisplayskip
 }%
 \def\Warning##1##2\\{%
   \def\linebreak{\hfill\break}%
   \hangindent=2.5em\hangafter=0
   \leavevmode
   {\scriptsize##1 : ##2}\par}%
}{%
 \par\smallskip
}

\usepackage{color}

\newenvironment{fshaded}{%
\MakeFramed {\FrameRestore}
}%
{\endMakeFramed}

\usepackage{tikz}
\usetikzlibrary{matrix}

\allowdisplaybreaks[4]

\begin{document}
\setlength{\baselineskip}{0.515cm}
\sloppy
\thispagestyle{empty}
\begin{flushleft}
DESY 16--148
\\
DO--TH 16/15  \\
September 2016\\
\end{flushleft}

\setcounter{table}{0}

\mbox{}
\vspace*{\fill}
\begin{center}

{\Large\bf The Asymptotic 3-Loop Heavy Flavor Corrections to the}

\vspace*{4mm}
{\Large \bf Charged Current  Structure Functions} 

\vspace*{4mm}
{\Large \bf \boldmath $F_L^{W^+-W^-}(x,Q^2)$ and $F_2^{W^+-W^-}(x,Q^2)$}

\vspace{4cm}
\large
A.~Behring$^a$,
J.~Bl\"umlein$^a$,
G.~Falcioni$^a$,
A.~De Freitas$^a$,
A.~von~Manteuffel$^b$,
and~C.~Schneider$^c$

\vspace{1.5cm}
\normalsize   
{\it  $^a$Deutsches Elektronen--Synchrotron, DESY,}\\
{\it  Platanenallee 6, D-15738 Zeuthen, Germany}

\vspace*{2mm}
{\it  $^b$Department of Physics and Astronomy, Michigan State University, \\ East Lansing, MI 48824, USA,\\
PRISMA Cluster of Excellence, Johannes Gutenberg University, 55099 Mainz, Germany}
\\

\vspace*{2mm}
{\it  $^c$Research Institute for Symbolic Computation (RISC),}\\
{\it Johannes Kepler University,
Altenbergerstra\ss{}e 69, A-4040 Linz, Austria}

\end{center}
\normalsize
\vspace{\fill}
\begin{abstract}
\noindent 
We calculate the massive Wilson coefficients for the heavy flavor contributions to the non-singlet charged
current deep-inelastic scattering structure functions $F_L^{W^+}(x,Q^2)-F_L^{W^-}(x,Q^2)$ and 
$F_2^{W^+}(x,Q^2)-F_2^{W^-}(x,Q^2)$ in the asymptotic region $Q^2 \gg m^2$ to 3-loop order in Quantum  Chromodynamics 
(QCD) at general values of the Mellin variable $N$ and the momentum fraction $x$. Besides the heavy quark pair 
production, also the single heavy flavor excitation $s \rightarrow c$ contributes. Numerical results are presented 
for the charm quark contributions and consequences on the unpolarized Bjorken sum rule and Adler sum rule are 
discussed.  
\end{abstract}

\vspace*{\fill}
\noindent
\numberwithin{equation}{section}

\newpage
\section{Introduction}
\label{sec:1}

\vspace*{1mm}
\noindent
The flavor non-singlet charged current structure functions $F_{1,2}^{W^+ - W^-}(x,Q^2)$ can be measured
in deep-inelastic neutrino(antineutrino)-nucleon scattering and in high energy charged lepton-nucleon scattering
in $e p$ or $\mu p$ collisions. They are associated with the well-known unpolarized Bjorken sum rule 
\cite{Bjorken:1967px} and Adler sum rule \cite{Adler:1965ty} by their first moment, the former of which can be 
used for QCD tests measuring the strong coupling constant $a_s = \alpha_s/(4\pi) = g_s^2/(4\pi)^2$. These 
structure functions also allow for an associated determination of the valence quark distributions 
of the nucleon. The massless contributions to these combinations of structure functions have been calculated recently
to 3-loop order \cite{Davies:2016ruz}. In the present paper we compute the asymptotic heavy flavor corrections to these
flavor non-singlet structure functions in the region $Q^2 \gg m^2$ to the same order, with $m$ the heavy 
quark mass and $Q^2$ the virtuality of the process,
and present numerical results in the case of charm quark contributions.

The massless and massive QCD corrections at first order in the coupling constant have been computed in
Refs.~\cite{Bardeen:1978yd,Furmanski:1981cw,Gluck:1997sj,Blumlein:2011zu}\footnote{The massive 1-loop corrections 
given in \cite{Gottschalk:1980rv} were corrected in \cite{Gluck:1997sj}, see also \cite{Blumlein:2011zu}.} and in
Refs.~\cite{Zijlstra:1992kj,Moch:1999eb,Buza:1997mg,Blumlein:2014fqa,Hasselhuhn:2013swa} to
$O(a_s^2)$\footnote{Some results given in \cite{Buza:1997mg} have been corrected in Ref.~\cite{Blumlein:2014fqa}.}.
The massive $O(a_s^2)$ corrections were calculated in the asymptotic representation \cite{Buza:1995ie}, which is valid 
at high scales $Q^2$. To obtain an estimate of the range of validity, one may perform an $O(a_s)$ comparison 
with the complete result for the process of single heavy quark excitation 
\cite{Gluck:1997sj,Blumlein:2011zu}. Likewise, a comparison is possible for the $O(a_s^2)$ corrections, 
which were 
given in complete form in Ref.~\cite{Blumlein:2016xcy} for the Wilson coefficient with the gauge boson coupling 
to the massless fermions and  assuming an approximation for the Cabibbo-suppressed flavor excitation term $s'  
\rightarrow c$, where the additional charm quark in the final state has been dealt with as being massless. 

The charged current scattering cross sections are given by
\cite{Arbuzov:1995id,Blumlein:2012bf}
\begin{eqnarray}
\frac{d \sigma^{\nu(\bar{\nu})}}{dx dy}
&=& \frac{G_F^2 s}{4 \pi} \frac{M_W^4}{(M_W^2 + Q^2)^2}
\\ && \times
\Biggl\{
\left(1 + (1-y)^2\right) F_2^{W^\pm}(x,Q^2)
- y^2 F_L^{W^\pm}(x,Q^2)
\pm \left(1 - (1-y)^2\right) xF_3^{W^\pm}(x,Q^2)\Biggr\}
\nonumber\\
\frac{d \sigma^{l(\bar{l})}}{dx dy}
&=& \frac{G_F^2 s }{4 \pi} \frac{M_W^4}{(M_W^2 + Q^2)^2}
\\ && \times
\Biggl\{
\left(1 + (1-y)^2\right) F_2^{W^\mp}(x,Q^2)
- y^2 F_L^{W^\mp}(x,Q^2)
\pm \left(1 - (1-y)^2\right) xF_3^{W^\mp}(x,Q^2)\Biggr\}~,
\nonumber
\end{eqnarray}
where $x = Q^2/ys$ and $y = q.P/l.P$ denote the Bjorken variables, $l$ and $P$ are the incoming lepton and nucleon 4-momenta,
and $s = (l+P)^2$. $G_F$ is the Fermi constant and $M_W$
the mass of the $W$-boson. $F_i^{W^\pm}(x,Q^2)$ are the structure functions, where the $+(-)$ signs refer to incoming neutrinos
(antineutrinos) and charged antileptons (leptons), respectively. We will consider the combination of structure functions
\begin{eqnarray}
\label{eq:comb}
F_{1,2}^{W^+ - W^-}(x,Q^2) = F_{1,2}^{W^+}(x,Q^2) - F_{1,2}^{W^-}(x,Q^2)
\end{eqnarray}
in the following. The longitudinal structure function is obtained by
\begin{eqnarray}
\label{eq:FL}
F_{L}(x,Q^2) = F_{2}(x,Q^2) - 2xF_{1}(x,Q^2).
\end{eqnarray}
The combinations (\ref{eq:comb}) can be measured projecting onto the kinematic factor $Y_+ = 1 + (1-y)^2$ 
in the case of $F_2$ for the differential cross
sections at $x, Q^2 = \rm const.$ and by varying $s$ in addition, in the case of $F_L$.

The main formalism to obtain the massive Wilson coefficients in the asymptotic range
$Q^2 \gg m^2$, i.e. $L_{q,L,2}^{W^+ - W^-, \rm NS}$ and $H_{q,L,2}^{W^+ - W^-, \rm NS}$,
has been outlined in Refs.~\cite{Buza:1995ie,Bierenbaum:2009mv,Blumlein:2014fqa}. They are 
composed of the massive non-singlet operator matrix elements (OMEs) \cite{Ablinger:2014vwa} and the
massless Wilson coefficients \cite{Davies:2016ruz} up to 3-loop order.
The following representation of the structure functions is obtained
\begin{eqnarray}
 F_{L,2}^{W^+ - W^-}(x,Q^2) &=&
2x \Biggl\{\Bigl[
  |V_{du}|^2 (d - \overline{d})
+ |V_{su}|^2 (s - \overline{s})
- V_{u}      (u - \overline{u})\Bigr] \otimes \Bigl[
C_{q,L,2}^{W^+ - W^-, \rm NS} 
\nonumber\\
&&
+L_{q,L,2}^{W^+ - W^-, \rm NS} \Bigr] 
+ \Bigl[
  |V_{dc}|^2 (d - \overline{d})
+ |V_{sc}|^2 (s - \overline{s}) \Bigr] \otimes
H_{q,L,2}^{W^+ - W^-, \rm NS}\Biggr\},
\end{eqnarray}
with one massless Wilson coefficient $C_{q,L,2}^{W^+ - W^-, \rm NS}$ and two massive Wilson coefficients $L_{q,L,2}^{W^+ -
W^-, \rm NS}$, $H_{q,L,2}^{W^+ - W^-, \rm NS}$, see Sections \ref{sec:2} and \ref{sec:3}. The 
coefficients $V_{ij}$ are the 
Cabibbo-Kobayashi-Maskawa
(CKM) \cite{Cabibbo:1963yz,Kobayashi:1973fv}
matrix elements, where $V_u = |V_{du}|^2 + |V_{su}|^2$, and the present numerical values are \cite{PDG}
\begin{eqnarray}
|V_{du}| &=& 0.97425,~~~
|V_{su}| = 0.2253,~~~
|V_{dc}| = 0.225,~~~
|V_{sc}| = 0.986~,
\end{eqnarray}  
with
\begin{eqnarray}
u - \overline{u}  &=& u_v, \\
d - \overline{d}  &=& d_v, \\
s - \overline{s}  &\approx& 0~.
\end{eqnarray}  
In the following we will consider only the charm quark corrections with $m \equiv m_c$ the charm quark 
mass in the on-shell scheme. The transformation to the $\overline{\rm MS}$ scheme has been given
in Ref.~\cite{Ablinger:2014vwa}. We note that the 3-loop asymptotic charm  quark corrections to the 
combination of structure functions $xF_3^{W^+ + W^-}(x,Q^2)$ have been calculated in 
Ref.~\cite{Behring:2015roa} and related corrections to the twist-2 contributions of the polarized 
structure functions $g_{1,2}(x,Q^2)$ in Ref.~\cite{Behring:2015zaa}.

A series of asymptotic 3-loop heavy flavor Wilson coefficients have also been calculated for neutral 
current scattering along with the transition matrix elements in the variable flavor number scheme, 
see~Refs.~\cite{Ablinger:2016swq} for recent surveys. 

\section{\boldmath The Structure Function $F_L(x,Q^2)$}
\label{sec:2}

\vspace*{1mm}
\noindent
The massive Wilson coefficients depend on the logarithms
\begin{eqnarray}
L_M = \ln\left(\frac{m^2}{\mu^2}\right),~~~~~~~~~L_Q = \ln\left(\frac{Q^2}{\mu^2}\right)~.
\end{eqnarray}
Here $\mu$ denotes the factorization scale. For the Wilson coefficients in Mellin $N$ space we consider 
the following series in the strong coupling constant
\begin{eqnarray}
L_{q,2(L)}^{W^+ - W^-, \rm NS}  = \delta_{2,0} +
\sum_{k=1}^\infty a_s^k L_{q,2(L)}^{W^+ - W^-, {\rm NS},(k)}, \\  
H_{q,2(L)}^{W^+ - W^-, \rm NS}  = \delta_{2,0} +
\sum_{k=1}^\infty a_s^k H_{q,2(L)}^{W^+ - W^-, {\rm NS},(k)}, \\  
C_{q,2(L)}^{W^+ - W^-, \rm NS}  = \delta_{2,0} +
\sum_{k=1}^\infty a_s^k C_{q,2(L)}^{W^+ - W^-, {\rm NS},(k)}.  
\end{eqnarray}
In the following we drop the arguments of the nested harmonic sums \cite{HSUM} and harmonic polylogarithms 
\cite{Remiddi:1999ew}
by defining $S_{\vec{a}}(N) \equiv S_{\vec{a}}$ and $H_{\vec{b}}(x) \equiv H_{\vec{b}}$.

The 3-loop contributions to the Wilson coefficient $L_{q,L}^{W^+ - W^-, {\rm NS},(3)}$ in Mellin $N$ space 
are given by 
\begin{eqnarray}
        \lefteqn{L_{q,L}^{W^+-W^-,\text{NS},(3)} =}
        \nonumber \\ &&
                \textcolor{blue}{C_F T_F^2 \big(2 N_F+1\big)} \Biggl[
                        L_Q^2 \frac{64}{9 (N+1)}
                        -L_Q \Biggl(
                                \frac{64 \big(19 N^2+7 N-6\big)}{27 N (N+1)^2}
                                +\frac{128 S_1}{9 (N+1)}
                        \Biggr)
                \Biggr]
        \nonumber \\ &&
                +\textcolor{blue}{C_A C_F T_F} \Biggl[
                        -L_Q^2 \frac{352}{9 (N+1)}
                        +L_Q \Biggl(
                                \frac{1088}{9 (N+1)} S_1
                                +\frac{128}{3 (N+1)} S_3
                                +\frac{128}{3 (N+1)} S_{-3}
        \nonumber \\ &&
                                +\biggl(
                                        -\frac{256 \big(N^2+N+1\big)}{3 (N-1) (N+1) (N+2)}
                                        +\frac{256}{3 (N+1)} S_1
                                \biggr) S_{-2}
                                -\frac{256}{3 (N+1)} S_{-2,1}
        \nonumber \\ &&
                                +\frac{16 P_4}{27 (N-1) N^2 (N+1)^3 (N+2)}
                                -\frac{128}{N+1} \zeta_3
                        \Biggr)
                \Biggr]
        \nonumber \\ &&
                +\textcolor{blue}{C_F^2 T_F} \Biggl[
                        L_Q^2 \Biggl(
                                \frac{8 \big(3 N^2+3 N+2\big)}{N (N+1)^2}
                                -\frac{32}{N+1} S_1
                        \Biggr)
                        +L_M^2 \Biggl(
                                \frac{8 \big(3 N^2+3 N+2\big)}{3 N (N+1)^2}
        \nonumber \\ &&
                                -\frac{32}{3 (N+1)} S_1
                        \Biggr)
                        +L_M \Biggl(
                                \frac{8 P_1}{9 N^2 (N+1)^3}
                                -\frac{320}{9 (N+1)} S_1
                                +\frac{64}{3 (N+1)} S_2
                        \Biggr)
        \nonumber \\ &&
                        +L_Q \Biggl(
                                \frac{128}{3 (N+1)} S_1^2
                                -\frac{32 P_2}{9 (N-1) N^2 (N+1)^3 (N+2)}
                                -\frac{16 (N+10) (5 N+3)}{9 N (N+1)^2} S_1
        \nonumber \\ &&
                                -\frac{256}{3 (N+1)} S_3
                                +\biggl(
                                        \frac{512 \big(N^2+N+1\big)}{3 (N-1) (N+1) (N+2)}
                                        -\frac{512}{3 (N+1)} S_1
                                \biggr) S_{-2}
        \nonumber \\ &&
                                -\frac{128}{3 (N+1)} S_2
                                -\frac{256}{3 (N+1)} S_{-3}
                                +\frac{512}{3 (N+1)} S_{-2,1}
                                +\frac{256}{N+1} \zeta_3
                        \Biggr)
        \nonumber \\ &&
                        +\frac{2 P_3}{27 N^3 (N+1)^4}
                        -\frac{896}{27 (N+1)} S_1
                        +\frac{160}{9 (N+1)} S_2
                        -\frac{32}{3 (N+1)} S_3
                \Biggr]+\hat{c}_{q,L}^{(3)},
\end{eqnarray}

where $\hat{c}_{q,L}^{(3)}=c^{(3)}_{q,L}(N_F+1)-c^{(3)}_{q,L}(N_F)$ is the 3-loop massless 
contribution, cf.~\cite{Davies:2016ruz}. The color factors in the case of QCD are $C_A= N_c = 3, C_F = 
(N_c^2-1)/(2N_c), T_F = 1/2$, $N_c = 3$ and $N_F$ denotes the number of massless flavors. Except for 
$\hat{c}^{(3)}_{q,L}(N_F)$, the Wilson coefficient is expressed by harmonic sums up to weight {\sf w=3}.
The polynomials $P_i$ above read
\begin{eqnarray}
        P_1 &=& 3 N^4+6 N^3+47 N^2+20 N-12 \\
        P_2 &=& 36 N^6+81 N^5-125 N^4-319 N^3-211 N^2-14 N-24 \\
        P_3 &=& 219 N^6+657 N^5+1193 N^4+763 N^3-40 N^2-48 N+72 \\
        P_4 &=& 469 N^6+1143 N^5-515 N^4-2055 N^3-746 N^2+120 N-144. 
\end{eqnarray}


By performing a Mellin inversion, the corresponding representation in $x$ space is obtained, which reads
\begin{eqnarray}
        \lefteqn{L_{q,L}^{W^+-W^-,\text{NS},(3)} =}
        \nonumber \\ &&
                \textcolor{blue}{C_A C_F T_F} \Biggl[
                        -L_Q^2 \frac{352 x}{9}
                        +L_Q \Biggl(
                                \frac{16}{27} (781 x-312)
                                +\biggl(
                                        \frac{128 \big(2 x^3+x^2-1\big) H_0}{3 x}
        \nonumber \\ &&
                                        +\frac{64}{3} x H_0^2
                                \biggr) H_{-1}
                                +\frac{832}{9} x H_0
                                -\frac{128}{3} x H_{-1}^2 H_0
                                +\biggl(
                                        \frac{1088 x}{9}
                                        +\frac{64}{3} x H_0^2
                                \biggr) H_1
        \nonumber \\ &&
                                -\frac{128}{3} x^2 H_0^2
                                +\biggl(
                                        -\frac{128 \big(2 x^3+x^2-1\big)}{3 x}
                                        +\frac{256}{3} x H_{-1}
                                        +\frac{128}{3} x H_0
                                \biggr) H_{0,-1}
        \nonumber \\ &&
                                -\frac{128}{3} x H_0 H_{0,1}
                                -\frac{256}{3} x H_{0,-1,-1}
                                -128 x H_{0,0,-1}
                                +\frac{128}{3} x H_{0,0,1}
                                +\biggl(
                                        \frac{256 x^2}{3}
        \nonumber \\ &&
                                        -\frac{128}{3} x H_{-1}
                                        -\frac{128}{3} x H_1
                                \biggr) \zeta_2
                        \Biggl)
                \Biggr]
        \nonumber \\ &&
                +\textcolor{blue}{C_F T_F^2 N_F} \Biggl[
                        L_Q^2 \frac{128 x}{9}
                        -L_Q \Biggl(
                                \frac{128}{27} (25 x-6)
                                +\frac{512}{9} x H_0
                                +\frac{256}{9} x H_1
                        \Biggr)
                \Biggr]
        \nonumber \\ &&
                +\textcolor{blue}{C_F T_F^2} \Biggl[
                        L_Q^2 \frac{64 x}{9}
                        -L_Q \Biggl(
                                \frac{64}{27} (25 x-6)
                                +\frac{256}{9} x H_0
                                +\frac{128}{9} x H_1
                        \Biggr)
                \Biggr]
        \nonumber \\ &&
                +\textcolor{blue}{C_F^2 T_F} \Biggl[
                        L_Q^2 \Biggl(
                                8 (x+2)
                                -16 x H_0
                                -32 x H_1
                        \Biggr)
                        +L_M^2 \Biggl(
                                \frac{8 (x+2)}{3}
                                -\frac{16}{3} x H_0
        \nonumber \\ &&
                                -\frac{32}{3} x H_1
                        \Biggr)
                        +L_M \Biggl(
                                \frac{32}{9} (x+3) H_0
                                -\frac{8}{9} (53 x-56)
                                -\frac{16}{3} x H_0^2
                                +\frac{64}{3} x H_{0,1}
        \nonumber \\ &&
                                +\biggl(
                                        -\frac{320 x}{9}
                                        -\frac{64}{3} x H_0
                                \biggr) H_1
                        \Biggr)
                        +L_Q \Biggl(
                                -\frac{32}{9} (83 x-47)
                                +\biggl(
                                        \frac{256}{3} x H_1
                                        +\frac{256}{3} x H_{-1}
        \nonumber \\ &&
                                        -\frac{32}{3} x (16 x+11)
                                \biggr) \zeta_2
                                -\biggl(
                                        \frac{256 \big(2 x^3+x^2-1\big) H_0}{3 x}
                                        +\frac{128}{3} x H_0^2
                                \biggr) H_{-1}
        \nonumber \\ &&
                                +\frac{256}{3} x H_{-1}^2 H_0
                                +\frac{16}{9} (89 x-24) H_0
                                +\frac{16}{3} x (16 x+7) H_0^2
                                +\biggl(
                                        \frac{80}{9} (5 x-6)
        \nonumber \\ &&
                                        +\frac{256}{3} x H_0
                                        -\frac{128}{3} x H_0^2
                                \biggr) H_1
                                +\frac{128}{3} x H_1^2
                                +\biggl(
                                        \frac{256 \big(2 x^3+x^2-1\big)}{3 x}
                                        -\frac{512}{3} x H_{-1}
        \nonumber \\ &&
                                        -\frac{256}{3} x H_0
                                \biggr) H_{0,-1}
                                +\biggl(
                                        32 x
                                        +\frac{256}{3} x H_0
                                \biggr) H_{0,1}
                                +\frac{512}{3} x H_{0,-1,-1}
                                +256 x H_{0,0,-1}
        \nonumber \\ &&
                                -\frac{256}{3} x H_{0,0,1}
                        \Biggr)
                        -\frac{2}{27} (653 x-872)
                        +\frac{16}{27} (11 x+42) H_0
                        +\frac{8}{9} (x+3) H_0^2
        \nonumber \\ &&
                        -\frac{8}{9} x H_0^3
                        +\biggl(
                                -\frac{896 x}{27}
                                -\frac{160}{9} x H_0
                                -\frac{16}{3} x H_0^2
                        \biggr) H_1
                        +\biggl(
                                \frac{160 x}{9}
                                +\frac{32}{3} x H_0
                        \biggr) H_{0,1}
        \nonumber \\ &&
                        -\frac{32}{3} x H_{0,0,1}
                \Biggr]+\hat{c}_{q,L}^{(3)}~.
\end{eqnarray}

Here $\zeta_k, k \in \mathbb{N}, k \geq 2$, are the values of the Riemann $\zeta$ function at integer 
argument.
Except for $\hat{c}^{(3)}_{q,L}(N_F)$, the Wilson coefficient is
expressed by weighted harmonic polylogarithms of up to weight {\sf w=3}.

The contribution of the massive Wilson coefficient $H_{q,L}$ is found by combining the massless Wilson 
coefficient $C_{q,L}$ and $L_{q,L}$:
\begin{equation}
H_{q,L}(x,Q^2) = C_{q,L}(N_F,x,Q^2) + L_{q,L}(x,Q^2)~.
\end{equation}
Eq.~(\ref{eq:FL}) provides the relation to the Wilson coefficients of the structure function 
$F_1(x,Q^2)$.

\section{\boldmath The Structure Function $F_2(x,Q^2)$}
\label{sec:3}

\vspace*{1mm}
\noindent
The asymptotic massive 3-loop Wilson coefficient $L_{q,2}^{W^+ - W^-, {\rm NS}, (3)}$ in Mellin $N$ space 
reads
\begin{eqnarray}
\lefteqn{L_{q,2}^{W^+-W^-,\text{NS},(3)}=}
        \nonumber \\ &&
        \textcolor{blue}{C_F T_F^2} \Biggl\{
        -\frac{2008 \big(3 N^2+3 N+2\big)}{243 N (N+1)}
        +L_M^3 \bigg(
                \frac{8 \big(3 N^2+3 N+2\big)}{9 N (N+1)}
                -\frac{32}{9} S_1
        \bigg)\nonumber\\ &&
        +L_M^2 \bigg(
                \frac{8 P_1}{27 N^2 (N+1)^2}
                -\frac{320}{27} S_1
                +\frac{64}{9} S_2
        \bigg)
        +L_M \bigg(
                \frac{2 P_3}{81 N^3 (N+1)^3}
                -\frac{896}{81} S_1\nonumber\\ &&
                +\frac{160}{27} S_2
                -\frac{32}{9} S_3
        \bigg)
        +\bigg(
                \frac{8032}{243}
                -\frac{128 \zeta_3}{3}
        \bigg) S_1
        +\frac{32 \big(3 N^2+3 N+2\big) \zeta_3}{3 N (N+1)}\nonumber\\ &&
        +\bigg(\textcolor{blue}{N_F}+\frac{1}{2}\bigg)\bigg[
		\frac{4 P_{41}}{729 N^4 (N+1)^4}
                +L_Q^3 \bigg(
                        \frac{32 \big(3 N^2+3 N+2\big)}{27 N (N+1)}
                        -\frac{128}{27} S_1
                \bigg)\nonumber\\ &&
                +L_M^3 \bigg(
                        \frac{16 \big(3 N^2+3 N+2\big)}{27 N (N+1)}
                        -\frac{64}{27} S_1
                \bigg)
                +L_Q^2 \bigg(
                        \frac{64}{9} S_1^2
                        -\frac{64}{3} S_2
                        -\frac{16 P_{11}}{27 N^2 (N+1)^2}\nonumber\\ &&
                        +\frac{32 \big(29 N^2+29 N-6\big) S_1}{27 N (N+1)}
                \bigg)
                +L_M \bigg(
                        \frac{4 P_{33}}{81 N^3 (N+1)^3}
                        -\frac{2176}{81} S_1
                        -\frac{320}{27} S_2\nonumber\\ &&
                        +\frac{64}{9} S_3
                \bigg)
                +L_Q \bigg(
                        \frac{16 P_{34}}{81 N^3 (N+1)^3}
                        +\big(
                                -\frac{32 P_{15}}{81 N^2 (N+1)^2}
                                +\frac{128}{9}S_2
                        \big) S_1
                        -\frac{128}{27} S_1^3\nonumber\\ &&
                        -\frac{32 \big(29 N^2+29 N-6\big) S_1^2}{27 N (N+1)}
                        +\frac{32 \big(35 N^2+35 N-2\big) S_2}{9 N (N+1)}
                        -\frac{1792}{27} S_3
                        +\frac{256}{9} S_{2,1}
                \bigg)\nonumber\\ &&
                +\bigg(
                        \frac{512 \zeta_3}{27}
                        -\frac{24064}{729}
                \bigg) S_1
                +\frac{128}{81} S_2
                +\frac{640}{81} S_3
                -\frac{128}{27} S_4
                -\frac{128 \big(3 N^2+3 N+2\big) \zeta_3}{27 N (N+1)}
        \bigg]
\Biggr\}\nonumber\\ &&
+\textcolor{blue}{C_F^2 T_F} \Biggl\{
        \frac{P_{47}}{162 N^5 (N+1)^5}
        -\frac{S_{-2,1}}{N^2 (N+1)} \frac{128}{81} \big(112 N^3+112 N^2-39 N+18\big)\nonumber\\ &&
        +L_Q^3 \bigg(
                \frac{1}{N^2 (N+1)^2} \frac{2}{3} \big(3 N^2+3 N+2\big)^2
                -\frac{16 \big(3 N^2+3 N+2\big) S_1}{3 N (N+1)}
                +\frac{32}{3} S_1^2
        \bigg)\nonumber\\ &&
        +L_Q^2 \bigg(
                -\frac{2 P_{31}}{9 N^3 (N+1)^3}
                +\big(
                        \frac{2 P_{16}}{9 N^2 (N+1)^2}
                        +\frac{176}{3} S_2
                \big) S_1
                -16 S_1^3
                +\frac{64}{3} S_{-3}
                +\frac{64}{3} S_3\nonumber\\ &&
                -\frac{4 \big(107 N^2+107 N-54\big) S_1^2}{9 N (N+1)}
                -\frac{44 \big(3 N^2+3 N+2\big) S_2}{3 N (N+1)}
                -\frac{128}{3} S_{-2,1}
                +\bigg(
                        \frac{128}{3} S_1\nonumber\\ &&
                        -\frac{64}{3 N (N+1)}
                \bigg) S_{-2}
        \bigg)
        +L_M^2 \bigg(
                L_Q \bigg(
                        \frac{1}{N^2 (N+1)^2} \frac{2}{3} \big(3 N^2+3 N+2\big)^2\nonumber\\ &&
                        -\frac{16 \big(3 N^2+3 N+2\big) S_1}{3 N (N+1)}
                        +\frac{32}{3} S_1^2
                \bigg)
		-\frac{2 (N-1) P_{25}}{3 N^3 (N+1)^3}
                +\bigg(
                        \frac{2 P_9}{3 N^2 (N+1)^2}
                        +\frac{80}{3} S_2
                \bigg) S_1\nonumber\\ &&
                -\frac{4 (N-1) (N+2) S_1^2}{N (N+1)}
                -\frac{16}{3} S_1^3
                -\frac{20 \big(3 N^2+3 N+2\big) S_2}{3 N (N+1)}
                +\frac{64}{3} S_3
                +\bigg(
                        \frac{128}{3} S_1\nonumber\\ &&
                        -\frac{64}{3 N (N+1)}
                \big) S_{-2}
                +\frac{64}{3} S_{-3}
                -\frac{128}{3} S_{-2,1}
        \bigg)
        +L_M \bigg(
                L_Q \bigg(
                        \bigg(
                                -\frac{8 P_{10}}{9 N^2 (N+1)^2}\nonumber\\ &&
                                -\frac{64}{3} S_2
                        \bigg) S_1
                        +\frac{1}{N^3 (N+1)^3} \frac{2}{9} \big(3 N^2+3 N+2\big) P_1
                        +\frac{320}{9} S_1^2\nonumber\\ &&
                        +\frac{16 \big(3 N^2+3 N+2\big) S_2}{3 N (N+1)}
                \bigg)
                +\frac{P_{40}}{9 N^4 (N+1)^4}
                -\bigg(
                        \frac{256}{3} S_3
                        +\frac{256}{3} S_{-2,1}
                        -64 \zeta_3\nonumber\\ &&
                        -\frac{2 P_{32}}{9 N^3 (N+1)^3}
                        -\frac{16 \big(59 N^2+59 N-6\big) S_2}{9 N (N+1)}
                \bigg) S_1
                +\bigg(
                        \frac{32}{3} S_2
                        -\frac{4 P_7}{3 N^2 (N+1)^2}
                \bigg) S_1^2\nonumber\\ &&
                -\frac{160}{9} S_1^3
                -\frac{4 P_{13} S_2}{9 N^2 (N+1)^2}
                -32 S_2^2
                +\frac{32 \big(29 N^2+29 N+12\big) S_3}{9 N (N+1)}
                -\frac{256}{3} S_4\nonumber\\ &&
                +\bigg(
                        -\frac{64 \big(16 N^2+10 N-3\big)}{9 N^2 (N+1)^2}
                        +\frac{1280}{9} S_1
                        -\frac{128}{3} S_2
                \big) S_{-2}
                -\frac{128}{3} S_{-4}
                +\frac{128}{3} S_{3,1}\nonumber\\ &&
                +\big(
                        \frac{64 \big(10 N^2+10 N+3\big)}{9 N (N+1)}
                        -\frac{128}{3} S_1
                \bigg) S_{-3}
                -\frac{128 \big(10 N^2+10 N-3\big) S_{-2,1}}{9 N (N+1)}\nonumber\\ &&
                -\frac{128}{3} S_{-2,2}
                +\frac{512}{3} S_{-2,1,1}
                -\frac{16 \big(3 N^2+3 N+2\big) \zeta_3}{N (N+1)}
        \bigg)
        +L_Q \bigg[
                \frac{4 P_{44}}{27 N^4 (N+1)^4 (N+2)}\nonumber\\ &&
                +\bigg(
                        -\frac{4 P_{37}}{27 N^3 (N+1)^3}
                        +\frac{640}{9} S_3
                        +\frac{64}{3} S_{2,1}
                        -\frac{32 \big(67 N^2+67 N-21\big) S_2}{9 N (N+1)}
                        +\frac{512}{3} S_{-2,1}\nonumber\\ &&
                        +64 \zeta_3
                \bigg) S_1
                +\bigg(
                        \frac{2 P_{19}}{27 N^2 (N+1)^2}
                        -\frac{224}{3} S_2
                \bigg) S_1^2
                +\frac{32 (4 N-1) (4 N+5) S_1^3}{9 N (N+1)}
                +\frac{80}{9} S_1^4\nonumber\\ &&
                +\frac{2 P_{18} S_2}{9 N^2 (N+1)^2}
                +48 S_2^2
                -\frac{32 \big(53 N^2+77 N+4\big) S_3}{9 N (N+1)}
                +\frac{352}{3} S_4
                +64 S_{-2}^2
                +\frac{448}{3} S_{-4}\nonumber\\ &&
                +\bigg(
                        -\frac{64 P_{23}}{9 N^2 (N+1)^2 (N+2)}
                        -\frac{128 \big(10 N^2+22 N-9\big) S_1}{9 N (N+1)}
                        -\frac{256}{3} S_1^2
                        +\frac{256}{3} S_2
                \bigg) S_{-2}\nonumber\\ &&
                +\bigg(
                        \frac{256}{3} S_1
                        -\frac{64 \big(10 N^2+22 N+3\big)}{9 N (N+1)}
                \bigg) S_{-3}
                +64 S_{3,1}
                +\frac{16 \big(9 N^2+9 N-2\big) S_{2,1}}{3 N (N+1)}\nonumber\\ &&
                +\frac{128 \big(10 N^2+22 N-9\big) S_{-2,1}}{9 N (N+1)}
                -\frac{256}{3} S_{-3,1}
                -64 S_{2,1,1}
                -\frac{512}{3} S_{-2,1,1}\nonumber\\ &&
                -\frac{16 \big(9 N^2-7 N+6\big) \zeta_3}{N (N+1)}
        \bigg]
        -\frac{1}{N (N+1)} \frac{48}{5} \big(3 N^2+3 N+2\big) \zeta_2^2
        +\bigg(
                \frac{P_{42}}{162 N^4 (N+1)^4}\nonumber\\ &&
                +\frac{8 P_{20} S_2}{81 N^2 (N+1)^2}
                -\frac{64}{9} S_2^2
                -\frac{8 \big(347 N^2+347 N+54\big) S_3}{27 N (N+1)}
                +\frac{704}{9} S_4
                +\frac{128 S_{2,1}}{9 N (N+1)}\nonumber\\ &&
                -\frac{320}{9} S_{3,1}
                -\frac{256 \big(10 N^2+10 N-3\big) S_{-2,1}}{27 N (N+1)}
                -\frac{256}{9} S_{-2,2}
                +\frac{64}{3} S_{2,1,1}
                +\frac{1024}{9} S_{-2,1,1}\nonumber\\ &&
                +\frac{192 \zeta_2^2}{5}
                -\frac{1208 \zeta_3}{9}
        \bigg) S_1
        +\bigg(
                \frac{P_{29}}{9 N^3 (N+1)^3}
                +\frac{16 \big(5 N^2+5 N-4\big) S_2}{9 N (N+1)}
                +16 S_3\nonumber\\ &&
                -\frac{128}{9} S_{2,1}
                -\frac{256}{9} S_{-2,1}
        \bigg) S_1^2
        +\bigg(
                -\frac{16 P_8}{27 N^2 (N+1)^2}
                +\frac{128}{27} S_2
        \bigg) S_1^3
        +\frac{512}{9} S_5\nonumber\\ &&
        +\bigg(
                \frac{P_{28}}{81 N^3 (N+1)^3}
                +\frac{400}{27} S_3
                +\frac{256}{3} S_{-2,1}
                -\frac{64 \zeta_3}{3}
        \bigg) S_2
        +\frac{8 P_{21} S_3}{81 N^2 (N+1)^2}\nonumber\\ &&
        -\frac{32 \big(23 N^2+23 N-3\big) S_2^2}{27 N (N+1)}
        -\frac{176 \big(17 N^2+17 N+6\big) S_4}{27 N (N+1)}
        +\bigg(
                -\frac{64 P_{14}}{81 N^3 (N+1)^3}\nonumber\\ &&
                +\frac{128 P_{12} S_1}{81 N^2 (N+1)^2}
                -\frac{128 S_1^2}{9 N (N+1)}
                +\frac{256}{27} S_1^3
                -\frac{1280}{27} S_2
                +\frac{512}{27} S_3
                -\frac{512}{9} S_{2,1}
        \bigg) S_{-2}\nonumber\\ &&
        +\bigg(
                \frac{1}{N (N+1)^2} \frac{64}{81} \big(112 N^3+224 N^2+169 N+39\big)
                +\frac{128}{9} S_1^2
                +\frac{128}{9} S_2\nonumber\\ &&
                -\frac{128 \big(10 N^2+10 N+3\big) S_1}{27 N (N+1)}
        \bigg) S_{-3}
        +\bigg(
                -\frac{128 \big(10 N^2+10 N+3\big)}{27 N (N+1)}
                +\frac{256}{9} S_1
        \bigg) S_{-4}\nonumber\\ &&
        +\frac{256}{9} S_{-5}
        +\frac{16 P_5 S_{2,1}}{9 N^2 (N+1)^2}
        +\frac{256}{9} S_{2,3}
        -\frac{512}{9} S_{2,-3}
        -\frac{512}{9} S_{4,1}
        +\frac{512}{9} S_{-2,3}\nonumber\\ &&
        +\frac{16 \big(89 N^2+89 N+30\big) S_{3,1}}{27 N (N+1)}
        -\frac{128 \big(10 N^2+10 N-3\big) S_{-2,2}}{27 N (N+1)}
        +\frac{512}{9} S_{2,1,-2}\nonumber\\ &&
        -\frac{16 \big(3 N^2+3 N+2\big) S_{2,1,1}}{3 N (N+1)}
        +\frac{256}{9} S_{3,1,1}
        +\frac{512 \big(10 N^2+10 N-3\big) S_{-2,1,1}}{27 N (N+1)}\nonumber\\ &&
        +\frac{512}{9} S_{-2,2,1}
        -\frac{2048}{9} S_{-2,1,1,1}
        +\frac{2 P_{17} \zeta_3}{9 N^2 (N+1)^2}
\Biggr\}\nonumber\\ &&
+\textcolor{blue}{C_A C_F T_F} \Biggl\{
        \frac{P_{46}}{729 N^5 (N+1)^5}
        +\frac{S_{-2,1}}{N^2 (N+1)} \frac{64}{81} \big(112 N^3+112 N^2-39 N+18\big)\nonumber\\ &&
        +L_M^3 \bigg(
                -\frac{44 \big(3 N^2+3 N+2\big)}{27 N (N+1)}
                +\frac{176}{27} S_1
        \bigg)
        +L_Q^3 \bigg(
                -\frac{88 \big(3 N^2+3 N+2\big)}{27 N (N+1)}\nonumber\\ &&
                +\frac{352}{27} S_1
        \bigg)
        +L_M^2 \bigg(
                \frac{1}{N^3 (N+1)^3} \frac{2}{9} \big(3 N^2+3 N+2\big) P_6
                -\frac{184}{9} S_1
                -\frac{32}{3} S_3
                -\frac{32}{3} S_{-3}\nonumber\\ &&
                +\big(
                        \frac{32}{3 N (N+1)}
                        -\frac{64}{3} S_1
                \big) S_{-2}
                +\frac{64}{3} S_{-2,1}
        \bigg)
        +L_Q^2 \bigg(
                -\frac{176}{9} S_1^2
                +\frac{176}{3} S_2
                -\frac{32}{3} S_{-3}\nonumber\\ &&
                -\frac{32}{3} S_3
                +\frac{2 P_{36}}{27 N^3 (N+1)^3}
                -\frac{16 \big(194 N^2+194 N-33\big) S_1}{27 N (N+1)}
                +\frac{64}{3} S_{-2,1}\nonumber\\ &&
                +\big(
                        \frac{32}{3 N (N+1)}
                        -\frac{64}{3} S_1
                \big) S_{-2}
        \bigg)
        +L_Q \bigg(
                -\frac{4 P_{45}}{81 N^4 (N+1)^4 (N+2)}
                +\frac{352}{27} S_1^3\nonumber\\ &&
                -\frac{S_2}{N (N+1)^2} \frac{16}{9} \big(230 N^3+460 N^2+213 N-11\big)
                +\bigg(
                        \frac{4 P_{38}}{81 N^3 (N+1)^3}
                        +32 S_3\nonumber\\ &&
                        -\frac{32 \big(11 N^2+11 N+3\big) S_2}{9 N (N+1)}
                        -\frac{128}{3} S_{2,1}
                        -\frac{256}{3} S_{-2,1}
                        -64 \zeta_3
                \bigg) S_1
                +\bigg(
                        \frac{32}{3} S_2\nonumber\\ &&
                        +\frac{16 \big(194 N^2+194 N-33\big)}{27 N (N+1)}
                \bigg) S_1^2
                -\frac{32}{3} S_2^2
                +\frac{16 \big(368 N^2+440 N-45\big) S_3}{27 N (N+1)}
                -\frac{224}{3} S_4\nonumber\\ &&
                +\bigg(
                        \frac{32 P_{23}}{9 N^2 (N+1)^2 (N+2)}
                        +\frac{64 \big(10 N^2+22 N-9\big) S_1}{9 N (N+1)}
                        +\frac{128}{3} S_1^2
                        -\frac{128}{3} S_2
                \bigg) S_{-2}\nonumber\\ &&
                -32 S_{-2}^2
                +\bigg(
                        \frac{32 \big(10 N^2+22 N+3\big)}{9 N (N+1)}
                        -\frac{128}{3} S_1
                \bigg) S_{-3}
                -\frac{224}{3} S_{-4}
                +\frac{128}{3} S_{-3,1}
                -\frac{64}{3} S_{3,1}\nonumber\\ &&
                -\frac{64 \big(11 N^2+11 N-3\big) S_{2,1}}{9 N (N+1)}
                -\frac{64 \big(10 N^2+22 N-9\big) S_{-2,1}}{9 N (N+1)}
                +64 S_{2,1,1}\nonumber\\ &&
                +\frac{256}{3} S_{-2,1,1}
                +\frac{32 \big(3 N^2-N+2\big) \zeta_3}{N (N+1)}
        \bigg)
        +L_M \bigg(
                \frac{P_{39}}{81 N^4 (N+1)^4}
                +\frac{1792}{27} S_2\nonumber\\ &&
                +\bigg(
                        -\frac{8 P_{30}}{81 N^3 (N+1)^3}
                        +32 S_3
                        +\frac{128}{3} S_{-2,1}
                        -64 \zeta_3
                \bigg) S_1
                -\frac{16 \big(31 N^2+31 N+9\big) S_3}{9 N (N+1)}\nonumber\\ &&
                +\frac{160}{3} S_4
                +\bigg(
                        \frac{32 \big(16 N^2+10 N-3\big)}{9 N^2 (N+1)^2}
                        -\frac{640}{9} S_1
                        +\frac{64}{3} S_2
                \bigg) S_{-2}
		-\frac{128}{3} S_{3,1}
                +\frac{64}{3} S_{-4}\nonumber\\ &&
                +\bigg(
                        -\frac{32 \big(10 N^2+10 N+3\big)}{9 N (N+1)}
                        +\frac{64}{3} S_1
                \bigg) S_{-3}
                +\frac{64 \big(10 N^2+10 N-3\big) S_{-2,1}}{9 N (N+1)}\nonumber\\ &&
                +\frac{64}{3} S_{-2,2}
                -\frac{256}{3} S_{-2,1,1}
                +\frac{16 \big(3 N^2+3 N+2\big) \zeta_3}{N (N+1)}
        \bigg)
        +\bigg(
                \frac{4 P_{43}}{729 N^4 (N+1)^4}\nonumber\\ &&
                -\frac{S_2}{N^2 (N+1)^2} \frac{16}{9} (N-1) \big(2 N^3-N^2-N-2\big)
                +\frac{112}{9} S_2^2
                +\frac{80 (2 N+1)^2 S_3}{9 N (N+1)}\nonumber\\ &&
                +\frac{64}{3} S_{3,1}
                -\frac{208}{9}S_4
                -\frac{8 \big(9 N^2+9 N+16\big) S_{2,1}}{9 N (N+1)}
                +\frac{128 \big(10 N^2+10 N-3\big) S_{-2,1}}{27 N (N+1)}\nonumber\\ &&
                +\frac{128}{9} S_{-2,2}
                -32 S_{2,1,1}
                -\frac{512}{9} S_{-2,1,1}
                -\frac{192 \zeta_2^2}{5}
                +\frac{4 \big(593 N^2+593 N+108\big) \zeta_3}{27 N (N+1)}
        \bigg) S_1\nonumber\\ &&
        +\frac{1}{N (N+1)} \frac{48}{5} \big(3 N^2+3 N\nonumber+2\big) \zeta_2^2
        +\bigg(
                \frac{4 P_{22}}{9 N^3 (N+1)^3}
                +\frac{32 S_2}{9 N (N+1)}
                -\frac{80}{9} S_3\nonumber\\ &&
                +\frac{128}{9} S_{2,1}
                +\frac{128}{9} S_{-2,1}
                -16 \zeta_3
        \bigg) S_1^2
        +\bigg(
                \frac{4 P_{35}}{81 N^3 (N+1)^3}
                +\frac{496}{27} S_3
                -\frac{64}{3} S_{2,1}\nonumber\\ &&
                -\frac{128}{3} S_{-2,1}
                +16 \zeta_3
        \bigg) S_2
        -\frac{64}{27} S_1^3 S_2
        -\frac{4 \big(15 N^2+15 N+14\big) S_2^2}{9 N (N+1)}
        -\frac{8 P_{26} S_3}{81 N^2 (N+1)^2}\nonumber\\ &&
        +\frac{4 \big(443 N^2+443 N+78\big) S_4}{27 N (N+1)}
        -\frac{224}{9} S_5
        +\bigg(
                \frac{32 P_{14}}{81 N^3 (N+1)^3}
                +\frac{64 S_1^2}{9 N (N+1)}\nonumber\\ &&
                -\frac{64 P_{12} S_1}{81 N^2 (N+1)^2}
                -\frac{128}{27} S_1^3
                +\frac{640}{27} S_2
                -\frac{256}{27} S_3
                +\frac{256}{9} S_{2,1}
        \bigg) S_{-2}
        -\bigg(
                \frac{64}{9} S_1^2
                +\frac{64}{9} S_2\nonumber\\ &&
                + \frac{32}{81}\frac{\big(112 N^3+224 N^2+169 N+39\big)}{N (N+1)^2} 
                -\frac{64 \big(10 N^2+10 N+3\big) S_1}{27 N (N+1)}
        \bigg) S_{-3}
        -\frac{128}{9} S_{-5}\nonumber\\ &&
        +\bigg(
                \frac{64 \big(10 N^2+10 N+3\big)}{27 N (N+1)}
                -\frac{128}{9} S_1
        \bigg) S_{-4}
        -\frac{128}{3} S_{2,3}
        -\frac{8 P_{24} S_{2,1}}{9 N^2 (N+1)^2}
        +\frac{256}{9} S_{2,-3}\nonumber\\ &&
        -\frac{8 (13 N+4) (13 N+9) S_{3,1}}{27 N (N+1)}
        +\frac{256}{9} S_{4,1}
        +\frac{64 \big(10 N^2+10 N-3\big) S_{-2,2}}{27 N (N+1)}
        +\frac{64}{3} S_{2,2,1}\nonumber\\ &&
        -\frac{256}{9} S_{-2,3}
        +\frac{8 \big(3 N^2+3 N+2\big) S_{2,1,1}}{N (N+1)}
        -\frac{256}{9} S_{2,1,-2}
        -\frac{256}{9} S_{3,1,1}
        -\frac{256}{9} S_{-2,2,1}\nonumber\\ &&
        -\frac{256 \big(10 N^2+10 N-3\big) S_{-2,1,1}}{27 N (N+1)}
        +\frac{224}{9} S_{2,1,1,1}
        +\frac{1024}{9} S_{-2,1,1,1}
        +\frac{P_{27} \zeta_3}{27 N^2 (N+1)^2}
\Biggr\}\nonumber\\ &&
+ \hat{c}^{(3)}_{q,2},
\end{eqnarray}

where $\hat{c}^{(3)}_{q,2}=c^{(3)}_{q,2}(N_F+1)-c^{(3)}_{q,2}(N_F)$ is obtained from the 3-loop 
massless Wilson coefficient Ref.~\cite{Davies:2016ruz}. Except for $\hat{c}^{(3)}_{q,2}(N_F)$, the Wilson 
coefficient is expressed by harmonic sums up to weight {\sf w=5}. The polynomials in the equation above 
are defined as follows
\begin{eqnarray*}
P_5&=&7 N^4+14 N^3+3 N^2-4 N-4\\ 
P_6&=&17 N^4+34 N^3+29 N^2+12 N+24\\ 
P_7&=&19 N^4+38 N^3-9 N^2-20 N+4\\ 
P_8&=&28 N^4+56 N^3+28 N^2+2 N+1\\ 
P_9&=&33 N^4+38 N^3-15 N^2-60 N-28\\ 
P_{10}&=&33 N^4+66 N^3+97 N^2+40 N-12\\ 
P_{11}&=&57 N^4+72 N^3+29 N^2-22 N-24\\ 
P_{12}&=&112 N^4+224 N^3+121 N^2+9 N+9\\ 
P_{13}&=&141 N^4+198 N^3+169 N^2-32 N-84\\ 
P_{14}&=&181 N^4+266 N^3+82 N^2-3 N+18\\ 
P_{15}&=&235 N^4+596 N^3+319 N^2+66 N+72\\ 
P_{16}&=&501 N^4+750 N^3+325 N^2-188 N-204\\ 
P_{17}&=&561 N^4+1122 N^3+767 N^2+302 N+48\\ 
P_{18}&=&1131 N^4+1926 N^3+1019 N^2-64 N-276\\ 
P_{19}&=&1139 N^4+3286 N^3+1499 N^2+504 N+828\\ 
P_{20}&=&1199 N^4+2398 N^3+1181 N^2+18 N+90\\ 
P_{21}&=&1220 N^4+2251 N^3+1772 N^2+303 N-138\\ 
P_{22}&=&3 N^5+11 N^4+10 N^3+19 N^2+23 N+16\\ 
P_{23}&=&6 N^5-25 N^3-45 N^2-11 N+6\\ 
P_{24}&=&12 N^5+16 N^4+18 N^3-15 N^2-5 N-8\\ 
P_{25}&=&15 N^5+39 N^4+39 N^3-17 N^2-32 N-20\\ 
P_{26}&=&27 N^5+863 N^4+1573 N^3+1151 N^2+144 N-36\\ 
P_{27}&=&648 N^5-2103 N^4-4278 N^3-3505 N^2-682 N-432\\ 
P_{28}&=&-11145 N^6-30915 N^5-33923 N^4-11449 N^3+1960 N^2-1032 N-2088\\ 
P_{29}&=&-151 N^6-469 N^5-181 N^4+305 N^3+80 N^2-88 N-56\\ 
P_{30}&=&155 N^6+465 N^5+465 N^4+155 N^3+108 N^2+108 N+54\\ 
P_{31}&=&216 N^6+459 N^5+417 N^4-3 N^3-125 N^2-80 N+12\\ 
P_{32}&=&309 N^6+647 N^5+293 N^4-783 N^3-718 N^2+68 N+216\\ 
P_{33}&=&525 N^6+1575 N^5+1535 N^4+973 N^3+536 N^2+48 N-72\\ 
P_{34}&=&609 N^6+1029 N^5+613 N^4-37 N^3-74 N^2+300 N+216\\ 
P_{35}&=&868 N^6+2469 N^5+2487 N^4+940 N^3+171 N^2+207 N+144\\ 
P_{36}&=&1407 N^6+3297 N^5+2891 N^4+727 N^3-514 N^2-240 N+144\\ 
P_{37}&=&1770 N^6+4731 N^5+4483 N^4+749 N^3+55 N^2+1440 N+756\\ 
P_{38}&=&7531 N^6+26121 N^5+27447 N^4+8815 N^3+1110 N^2+936 N-324\\ 
P_{39}&=&-4785 N^8-19140 N^7-18754 N^6+1320 N^5+12723 N^4+6548 N^3+4080 N^2
        \nonumber \\ &&
        -648 N-1728\\ 
P_{40}&=&-45 N^8-138 N^7-774 N^6-476 N^5-881 N^4-762 N^3-868 N^2-88 N+192\\ 
P_{41}&=&3549 N^8+14196 N^7+23870 N^6+25380 N^5+15165 N^4+1712 N^3-2016 N^2
        \nonumber \\ &&
        +144 N+432\\ 
P_{42}&=&-3456 B_4 N^4 (N+1)^4+42591 N^8+161388 N^7+226848 N^6+105790 N^5
        \nonumber \\ &&
        -26735 N^4-28666 N^3+3560 N^2-3192 N-4464\\ 
P_{43}&=&1944 B_4 N^4 (N+1)^4-10807 N^8-43228 N^7-63222 N^6-40150 N^5-14587 N^4
        \nonumber \\ &&
        -9018 N^3-7452 N^2-2376 N-324\\ 
P_{44}&=&828 N^9+3456 N^8+4539 N^7+2412 N^6+1852 N^5+5026 N^4+4703 N^3+2468 N^2
        \nonumber \\ &&
        -324 N-576\\ 
P_{45}&=&8274 N^9+39795 N^8+71627 N^7+64189 N^6+29919 N^5+8096 N^4+5620 N^3
        \nonumber \\ &&
        +5664 N^2-1368 N-2160\\ 
P_{46}&=&-1944 B_4 N^4 (N+1)^4 (3 N^2+3 N+2)+165 N^{10}+825 N^9+109664 N^8+331682 N^7
        \nonumber \\ &&
        +457641 N^6+346145 N^5+219290 N^4+86724 N^3+13608 N^2+14256 N+10368\\ 
P_{47}&=&864 B_4 N^4 (N+1)^4 (3 N^2+3 N+2)-18351 N^{10}-87156 N^9-198195 N^8-244182 N^7
        \nonumber \\ &&
        -184797 N^6-70160 N^5-23209 N^4-8030 N^3-984 N^2-2328 N-2160~.
\end{eqnarray*}

Here the constant $B_4$ is given by
\begin{eqnarray}
B_4 = - 4 \zeta_2 \ln^2(2) + \frac{2}{3} \ln^4(2) - \frac{13}{2} \zeta_4 + 16 \Li_4\left(\frac{1}{2}\right)~.
\end{eqnarray}

By performing the Mellin inversion to $x$-space one obtains
\begin{eqnarray}
\lefteqn{L_{q,2}^{W^+-W^-,{\rm NS},(3)}=}
        \nonumber \\ &&
        \delta(1-x)\bigg\{
\textcolor{blue}{C_A C_F T_F} \bigg[
        -\frac{44 L_M^3}{9}
        -\frac{88 L_Q^3}{9}
        +L_M^2 \bigg(
                \frac{34}{3}
                -\frac{16 \zeta_3}{3}
        \bigg)
        +L_Q^2 \bigg(
                \frac{938}{9}\nonumber\\ &&
                -\frac{16 \zeta_3}{3}
        \bigg)
        +L_M \bigg(
                -\frac{1595}{27}
                +\frac{136 \zeta_2^2}{15}
                +\frac{272 \zeta_3}{9}
        \bigg)
        +L_Q \bigg(
                -\frac{11032}{27}
                -\frac{32 \zeta_2}{3}
                -\frac{392 \zeta_2^2}{15}\nonumber\\ &&
                +\frac{1024 \zeta_3}{9}
        \bigg)
        +\frac{5248 \zeta_2^2}{135}
        -\frac{10045 \zeta_3}{81}
        -\frac{16}{9} \zeta_2 \zeta_3
        -\frac{176 \zeta_5}{9}
        +\frac{55}{243}
        -8 B_4
\bigg]\nonumber\\ &&
+\textcolor{blue}{C_F^2 T_F} \bigg[
        6 L_M^2 L_Q
        +6 L_Q^3
        +L_M^2 \bigg(
                \frac{32 \zeta_3}{3}
                -10
        \bigg)
        +2 L_M L_Q
        +L_Q^2 \bigg(
                \frac{32 \zeta_3}{3}
                -48
        \bigg)\nonumber\\ &&
        -L_M \bigg(
                5
                +\frac{272 \zeta_2^2}{15}
                +\frac{112 \zeta_3}{9}
        \bigg)
        +L_Q \bigg(
                \frac{368}{3}
                +\frac{64 \zeta_2}{3}
                +\frac{784 \zeta_2^2}{15}
                -\frac{1616 \zeta_3}{9}
        \bigg)
        -\frac{6608 \zeta_2^2}{135}\nonumber\\ &&
        +\frac{13682 \zeta_3}{81}
        +\frac{32 \zeta_2 \zeta_3}{9}
        +\frac{352 \zeta_5}{9}
        -\frac{2039}{18}
        +16 B_4
\bigg]
+\textcolor{blue}{C_F N_F T_F^2} \bigg[
        \frac{16 L_M^3}{9}
        +\frac{32 L_Q^3}{9}\nonumber\\ &&
        -\frac{304 L_Q^2}{9}
        +\frac{700 L_M}{27}
        +\frac{3248 L_Q}{27}
        -\frac{128 \zeta_3}{9}
        +\frac{4732}{243}
\bigg]
+\textcolor{blue}{C_F T_F^2} \bigg[
        \frac{32 L_M^3}{9}
        +\frac{16 L_Q^3}{9}\nonumber\\ &&
        +\frac{8 L_M^2}{9}
        -\frac{152 L_Q^2}{9}
        +\frac{496 L_M}{27}
        +\frac{1624 L_Q}{27}
        +\frac{224 \zeta_3}{9}
        -\frac{3658}{243}
\bigg]
\bigg\}\nonumber\\ &&
+\bigg\{
\frac{\textcolor{blue}{C_F N_F T_F^2}}{1-x} \bigg[
        \frac{64 L_M^3}{27}
        +\frac{128 L_Q^3}{27}
        -L_Q^2 \bigg(
                \frac{928}{27}
                +\frac{256}{9} H_0
                +\frac{128}{9} H_1
        \bigg)
        +L_M \bigg(
                \frac{2176}{81}\nonumber\\ &&
                -\frac{320}{27} H_0
                -\frac{32}{9} H_0^2
        \bigg)
        +L_Q \bigg(
                \frac{7520}{81}
                +\frac{4288 H_0}{27}
                +\frac{128}{3} H_0^2
                +\frac{1856 H_1}{27}
                +\frac{256}{9} H_0 H_1\nonumber\\ &&
                +\frac{128}{9} H_1^2
                +\frac{256}{9} H_{0,1}
                -\frac{512 \zeta_2}{9}
        \bigg)
        +\frac{128 H_0}{81}
        -\frac{320}{81} H_0^2
        -\frac{64}{81} H_0^3
        -\frac{512 \zeta_3}{27}
        +\frac{24064}{729}
\bigg]\nonumber\\ &&
+\frac{\textcolor{blue}{C_F T_F^2}}{1-x} \bigg[
        \frac{128 L_M^3}{27}
        +\frac{64 L_Q^3}{27}
        +L_M^2 \bigg(
                \frac{320}{27}
                +\frac{64 H_0}{9}
        \bigg)
        -L_Q^2 \bigg(
                \frac{464}{27}
                +\frac{128}{9} H_0\nonumber\\ &&
                +\frac{64}{9} H_1
        \bigg)
        +\frac{1984 L_M}{81}
        +L_Q \bigg(
                \frac{3760}{81}
                +\frac{2144 H_0}{27}
                +\frac{64}{3} H_0^2
                +\frac{928 H_1}{27}
                +\frac{128}{9} H_0 H_1\nonumber\\ &&
                +\frac{64}{9} H_1^2
                +\frac{128}{9} H_{0,1}
                -\frac{256 \zeta_2}{9}
        \bigg)
        +\frac{64 H_0}{81}
        -\frac{160}{81} H_0^2
        -\frac{32}{81} H_0^3
        +\frac{896 \zeta_3}{27}
        -\frac{12064}{729}
\bigg]\nonumber\\ &&
+\frac{\textcolor{blue}{C_A C_F T_F}}{(1-x)^2} \bigg[
                        \frac{32}{9} (x+2) H_{0,1}
                	-\frac{4}{81} (800 x-773) H_0^2
                        +\frac{32}{81} (94 x-121) \zeta_2
        	\bigg]\nonumber\\ &&
+\frac{\textcolor{blue}{C_A C_F T_F}}{1-x} \bigg[
        -\frac{176 L_M^3}{27}
        -\frac{352 L_Q^3}{27}
        +L_Q^2 \bigg(
                \frac{3104}{27}
                +\frac{704 H_0}{9}
                +\frac{16}{3} H_0^2
                +\frac{352 H_1}{9}
                -\frac{32 \zeta_2}{3}
        \bigg)\nonumber\\ &&
        +L_M^2 \bigg(
                \frac{184}{9}
                +\frac{16}{3} H_0^2
                -\frac{32 \zeta_2}{3}
        \bigg)
        +L_Q \bigg(
                -\frac{30124}{81}
                -\frac{14144}{27} H_0
                -\frac{1216}{9} H_0^2
                -\frac{80}{9} H_0^3\nonumber\\ &&
                -\frac{6208}{27} H_1
                -\frac{704}{9} H_0 H_1
                -\frac{16}{3} H_0^2 H_1
                -\frac{352}{9} H_1^2
                +\frac{32}{3} H_0 H_1^2
                -64 H_0 H_{0,-1}\nonumber\\ &&
                -\frac{704}{9} H_{0,1}
                +\frac{32}{3} H_0 H_{0,1}
                -\frac{128}{3} H_1 H_{0,1}
                +128 H_{0,0,-1}
                -\frac{128}{3} H_{0,0,1}
                +64 H_{0,1,1}\nonumber\\ &&
                +\bigg(
                        192
                        +\frac{128 H_0}{3}
                        +64 H_1
                \bigg) \zeta_2
                -\frac{256 \zeta_3}{3}
        \bigg)
        +L_M \bigg(
                \frac{1240}{81}
                +\frac{1792 H_0}{27}
                +\frac{248}{9} H_0^2\nonumber\\ &&
                +\frac{32}{9} H_0^3
                -16 H_0^2 H_1
                +32 H_0 H_{0,1}
                -\frac{64}{3} H_{0,0,1}
                +\bigg(
                        -\frac{320}{9}
                        -\frac{64}{3} H_0
                \bigg) \zeta_2
                +96 \zeta_3
        \bigg)\nonumber\\ &&
        +\bigg(
                -\frac{496}{27} H_0
                -\frac{112}{9} H_0^2
                +8 H_1
                -\frac{160}{9} H_0 H_1
                -\frac{128}{9} H_1^2
                +\frac{32}{9} H_{0,1}
        \bigg) \zeta_2
        +\frac{43228}{729}\nonumber\\ &&
        -\frac{32 B_4}{3}
        +\frac{3256 H_0}{81}
        +\frac{496}{81} H_0^3
        +\frac{16}{27} H_0^4
        +\frac{32}{3} H_1
        -\frac{32}{9} H_0 H_1
        -\frac{160}{9} H_0^2 H_1\nonumber\\ &&
        -\frac{112}{27} H_0^3 H_1
        +\frac{8}{9} H_0^2 H_1^2
        -\frac{64}{27} H_0 H_1^3
        +\frac{368}{9} H_0 H_{0,1}
        +\frac{16}{3} H_0^2 H_{0,1}
        -8 H_1 H_{0,1}\nonumber\\ &&
        -\frac{128}{9} H_0 H_1 H_{0,1}
        +\frac{128}{9} H_1^2 H_{0,1}
        -\frac{32}{9} H_{0,1}^2
        -\frac{1072}{27} H_{0,0,1}
        +\frac{32}{9} H_0 H_{0,0,1}\nonumber\\ &&
        +\frac{320}{9} H_1 H_{0,0,1}
        +24 H_{0,1,1}
        +\frac{160}{9} H_0 H_{0,1,1}
        -32 H_1 H_{0,1,1}
        -\frac{224}{9} H_{0,0,1,1}\nonumber\\ &&
        +\frac{224}{9} H_{0,1,1,1}
        +\frac{592 \zeta_2^2}{15}
        +\bigg(
                -\frac{1196}{27}
                +\frac{160 H_0}{9}
                -\frac{32}{9} H_1
        \bigg) \zeta_3
\bigg]\nonumber\\ &&
+\frac{\textcolor{blue}{C_F^2 T_F}}{1-x} \bigg[
        L_M^2 L_Q \bigg(
                16
                -\frac{32}{3} H_0
                -\frac{64}{3} H_1
        \bigg)
        +L_Q^3 \bigg(
                16
                -\frac{32}{3} H_0
                -\frac{64}{3} H_1
        \bigg)
        -L_M^2 \bigg(
                22\nonumber\\ &&
                +16 H_0
                -\frac{16}{3} H_0^2
                -8 H_1
                -\frac{128}{3} H_0 H_1
                -16 H_1^2
                +\frac{64 \zeta_2}{3}
        \bigg)
        +L_Q^2 \bigg(
                -\frac{334}{3}
                +\frac{32 H_0}{9}\nonumber\\ &&
                +\frac{80}{3} H_0^2
                +\frac{856 H_1}{9}
                +\frac{320}{3} H_0 H_1
                +48 H_1^2
                -\frac{256 \zeta_2}{3}
        \bigg)
        +L_M L_Q \bigg(
                \frac{88}{3}
                -\frac{176}{9} H_0\nonumber\\ &&
                -\frac{32}{3} H_0^2
                -\frac{640}{9} H_1
                -\frac{64}{3} H_0 H_1
                +\frac{64 \zeta_2}{3}
        \bigg)
        +L_M \bigg(
                \frac{88}{9} H_0^2
                -\frac{206}{3}
                -\frac{112}{3} H_0
                +\frac{64}{9} H_0^3\nonumber\\ &&
                +\frac{152 H_1}{3}
                +\frac{1424}{9} H_0 H_1
                +\frac{160}{3} H_0^2 H_1
                +\frac{160}{3} H_1^2
                +\frac{32}{3} H_0 H_1^2
                -\frac{64}{3} H_0 H_{0,1}
                -\bigg(
                        \frac{784}{9}\nonumber\\ &&
                        +\frac{128}{3}H_0
                        +\frac{64}{3} H_1
                \bigg) \zeta_2
                -64 \zeta_3
        \bigg)
        +L_Q \bigg(
                \frac{2360}{9}
                +\frac{4508 H_0}{27}
                -\frac{160}{3} H_0^2
                -\frac{224}{9} H_0^3\nonumber\\ &&
                -\frac{4556}{27} H_1
                -\frac{3680}{9} H_0 H_1
                -128 H_0^2 H_1
                -\frac{512}{3} H_1^2
                -128 H_0 H_1^2
                -\frac{320}{9} H_1^3\nonumber\\ &&
                +128 H_0 H_{0,-1}
                +48 H_{0,1}
                -\frac{64}{3} H_0 H_{0,1}
                +\frac{64}{3} H_1 H_{0,1}
                -256 H_{0,0,-1}
                -64 H_{0,1,1}\nonumber\\ &&
                +\bigg(
                        \frac{2608}{9}
                        +\frac{832 H_0}{3}
                        +\frac{448 H_1}{3}
                \bigg) \zeta_2
                +320 \zeta_3
        \bigg)
        +\frac{64 B_4}{3}
        -\bigg(
                \frac{2488}{27}
                +\frac{1192}{27} H_0
                +\frac{80}{9} H_0^2\nonumber\\ &&
                +\frac{160}{9} H_1
                -\frac{32}{3} H_0 H_1
                -\frac{128}{9} H_1^2
                +\frac{32}{9} H_{0,1}
        \bigg) \zeta_2
        -\frac{3262}{27} H_0
        +\frac{196}{27} H_0^2
        +\frac{380}{81} H_0^3\nonumber\\ &&
        +\frac{4}{3} H_0^4
        +\frac{302 H_1}{9}
        +\frac{13624}{81} H_0 H_1
        +\frac{1628}
        {27} H_0^2 H_1
        +\frac{304}{27} H_0^3 H_1
        +\frac{448}{9} H_1^2\nonumber\\ &&
        +\frac{80}{9} H_0 H_1^2
        -\frac{8}{9} H_0^2 H_1^2
        +\frac{128}{27} H_0 H_1^3
        +\frac{112}{9} H_{0,1}
        -\frac{1304}{27} H_0 H_{0,1}
        -\frac{32}{3} H_0^2 H_{0,1}\nonumber\\ &&
        +\frac{160}{9} H_0 H_1 H_{0,1}
        -\frac{128}{9} H_1^2 H_{0,1}
        -\frac{16}{9} H_{0,1}^2
        +\frac{1184}{27} H_{0,0,1}
        +\frac{128}{9} H_0 H_{0,0,1}\nonumber\\ &&
        -\frac{256}{9} H_1 H_{0,0,1}
        -16 H_{0,1,1}
        -\frac{32}{3} H_0 H_{0,1,1}
        +\frac{64}{3} H_1 H_{0,1,1}
        -\frac{128}{9} H_{0,0,0,1}\nonumber\\ &&
        +\frac{64}{3} H_{0,0,1,1}
        -\frac{1328 \zeta_2^2}{45}
        +\big(
                \frac{3088}{27}
                -\frac{128}{9} H_0
                +\frac{160 H_1}{9}
        \big) \zeta_3
        -\frac{14197}{54}
\bigg]
\bigg\}_+\nonumber\\ &&
+\bigg\{
\textcolor{blue}{C_F^2 T_F} \bigg[
        L_M^2 L_Q \bigg(
                8 (x+1) H_0
                +\frac{32}{3} (x+1) H_1
                -\frac{8}{3} (x+5)
        \bigg)\nonumber\\ &&
        +L_Q^3 \bigg(
                8 (x+1) H_0
                +\frac{32}{3} (x+1) H_1
                -\frac{8}{3} (x+5)
        \bigg)
        +L_M L_Q \bigg(
                \frac{4}{9} (19 x-85)\nonumber\\ &&
                +\frac{8}{3} (13 x+1) H_0
                +8 (x+1) H_0^2
                +\frac{128}{9} (4 x+1) H_1
                +\frac{32}{3} (x+1) H_0 H_1\nonumber\\ &&
                -\frac{32}{3} (x+1) \zeta_2
        \bigg)
        +L_M^2 \bigg(
                12 (5 x-2)
                -16 (2 x+1) H_0
                +\frac{64 \big(x^2+1\big) H_{-1} H_0}{3 (x+1)}\nonumber\\ &&
                -\frac{4 \big(9 x^2+10 x+9\big) H_0^2}{3 (x+1)}
                -\frac{16}{3} (3 x+2) H_1
                -\frac{64}{3} (x+1) H_0 H_1
                -8 (x+1) H_1^2\nonumber\\ &&
                -\frac{64 \big(x^2+1\big) H_{0,-1}}{3 (x+1)}
                -\frac{8}{3} (x+1) H_{0,1}
                +\frac{8 \big(9 x^2+10 x+9\big) \zeta_2}{3 (x+1)}
        \bigg)\nonumber\\ &&
        +L_Q^2 \bigg(
                \frac{4}{9} (188 x+157)
                -\frac{88}{3} (3 x+1) H_0
                +\frac{64 \big(x^2+1\big) H_{-1} H_0}{3 (x+1)}
                -24 (x+1) H_1^2\nonumber\\ &&
                -\frac{16}{9} (59 x+26) H_1
                -\frac{4 \big(21 x^2+34 x+21\big) H_0^2}{3 (x+1)}
                -\frac{160}{3} (x+1) H_0 H_1\nonumber\\ &&
                -8 (x+1) H_{0,1}
                -\frac{64 \big(x^2+1\big) H_{0,-1}}{3 (x+1)}
                +\frac{8 \big(23 x^2+38 x+23\big) \zeta_2}{3 (x+1)}
        \bigg)\nonumber\\ &&
        +L_M \bigg(
                \frac{4}{3} (171 x-116)
                +\bigg(
                        \frac{8 \big(117 x^2+118 x+81\big)}{9 (x+1)}
                        +\frac{128 \big(x^2+1\big) H_{-1}}{3 (x+1)}\nonumber\\ &&
                        +\frac{16 (x+3) (3 x+1) H_0}{3 (x+1)}
                        +\frac{32}{3} (x+1) H_1
                \bigg) \zeta_2
                -\frac{4}{3} (107 x+89) H_0\nonumber\\ &&
                +\frac{256 \big(4 x^2+3 x+4\big) H_{-1} H_0}{9 (x+1)}
                -\frac{4 \big(201 x^2+250 x+129\big) H_0^2}{9 (x+1)}\nonumber\\ &&
                +\frac{32 \big(x^2+1\big) H_{-1} H_0^2}{3 (x+1)}
                -\frac{32 \big(3 x^2+4 x+3\big) H_0^3}{9 (x+1)}
                -\frac{4}{9} (327 x-73) H_1\nonumber\\ &&
                -\frac{32}{9} (38 x+17) H_0 H_1
                -\frac{80}{3} (x+1) H_0^2 H_1
                -\frac{16}{3} (7 x+3) H_1^2
                -\frac{16}{3} (x+1) H_0 H_1^2\nonumber\\ &&
                -\frac{256 \big(4 x^2+3 x+4\big) H_{0,-1}}{9 (x+1)}
                +\frac{280}{9} (x+1) H_{0,1}
                -\frac{128 \big(x^2+1\big) H_{-1} H_{0,1}}{3 (x+1)}\nonumber\\ &&
                +\frac{32}{3} (x+1) H_0 H_{0,1}
                +\frac{128 \big(x^2+1\big) H_{0,-1,1}}{3 (x+1)}
                -\frac{64 \big(x^2+1\big) H_{0,0,-1}}{3 (x+1)}\nonumber\\ &&
                +\frac{16 \big(3 x^2-2 x+3\big) H_{0,0,1}}{3 (x+1)}
                +\frac{128 \big(x^2+1\big) H_{0,1,-1}}{3 (x+1)}
                +\frac{16 \big(x^2+14 x+1\big) \zeta_3}{3 (x+1)}
        \bigg)\nonumber\\ &&
        +L_Q \bigg(
                -\frac{8}{27} (1925 x-284)
                -\frac{64}{9}\frac{\big(36 x^3+61 x^2+18 x+13\big)H_{-1} H_0}{x+1} \nonumber\\ && 
                +\frac{64}{9}\frac{\big(36 x^3+61 x^2+18 x+13\big)H_{0,-1}}{x+1} 
                +\frac{4}{9}\frac{\big(288 x^3+801 x^2+742 x+309\big)H_0^2}{x+1}\nonumber\\ &&  
                +\bigg(
                        -\frac{32}{9}\frac{\big(72 x^3+199 x^2+180 x+73\big)}{x+1}  
                        +\frac{64 \big(7 x^2+6 x+3\big) H_{-1}}{3 (x+1)}\nonumber\\ &&
                        -\frac{16 \big(35 x^2+66 x+35\big) H_0}{3 (x+1)}
                        +32 (x-3) H_1
                \bigg) \zeta_2
                +\frac{8 \big(186 x^2+211 x+73\big) H_0}{9 (x+1)}\nonumber\\ &&
                +\frac{64 \big(7 x^2+6 x+3\big) H_{-1}^2 H_0}{3 (x+1)}
                -\frac{32 \big(11 x^2+6 x+7\big) H_{-1} H_0^2}{3 (x+1)}
                +\frac{8}{27} (425 x+434) H_1\nonumber\\ &&
                +\frac{32 \big(9 x^2+13 x+9\big) H_0^3}{9 (x+1)}
                +\frac{16}{9} (193 x+121) H_0 H_1
                +\frac{32}{3} (x+7) H_0^2 H_1\nonumber\\ &&
                +\frac{32}{3} (15 x+8) H_1^2
                +64 (x+1) H_0 H_1^2
                +\frac{160}{9} (x+1) H_1^3
                +\frac{16}{9} (61 x+13) H_{0,1}\nonumber\\ &&
                -\frac{128 \big(7 x^2+6 x+3\big) H_{-1} H_{0,-1}}{3 (x+1)}
                -\frac{128 x (3 x+5) H_0 H_{0,-1}}{3 (x+1)}
                +\frac{16}{3} (25 x+1) H_0 H_{0,1}\nonumber\\ &&
                -\frac{32}{3} (x+1) H_1 H_{0,1}
                +\frac{128 \big(7 x^2+6 x+3\big) H_{0,-1,-1}}{3 (x+1)}
                +\frac{64 \big(23 x^2+26 x+7\big) H_{0,0,-1}}{3 (x+1)}\nonumber\\ &&
                -\frac{16}{3} (19 x-5) H_{0,0,1}
                +48 (x+1) H_{0,1,1}
                -\frac{32 \big(21 x^2+38 x+25\big) \zeta_3}{3 (x+1)}
        \bigg)\nonumber\\ &&
        +\bigg(
                \frac{4 \big(1619 x^2+1338 x+1511\big)}{81 (x+1)}
                +\frac{64 \big(29 x^2+18 x+29\big) H_{-1}}{27 (x+1)}
                +\frac{64 \big(x^2+1\big) H_{-1}^2}{9 (x+1)}\nonumber\\ &&
                +\frac{4 \big(147 x^2+298 x-9\big) H_0}{27 (x+1)}
                +\frac{64 \big(x^2+1\big) H_{-1} H_0}{9 (x+1)}
                +\frac{4 (x+5) (5 x+1) H_0^2}{9 (x+1)}\nonumber\\ &&
                +\frac{16}{9} (x+9) H_1
                -\frac{16}{3} (x+1) H_0 H_1
                -\frac{64}{9} (x+1) H_1^2
                +\frac{16}{9} (x+1) H_{0,1}
        \bigg) \zeta_2\nonumber\\ &&
        +\frac{1}{x+1} \frac{8}{45} \big(131 x^2+178 x+131\big) \zeta_2^2
        -\bigg(
                \frac{8 \big(235 x^2+404 x+409\big)}{27 (x+1)}
                +\frac{80}{9} (x+1) H_1\nonumber\\ &&
                +\frac{256 \big(x^2+1\big) H_{-1}}{9 (x+1)}
                -\frac{8 \big(15 x^2+22 x+15\big) H_0}{9 (x+1)}
        \bigg) \zeta_3
        -\frac{32}{3} B_4 (x+1)\nonumber\\ &&
        +\frac{11}{27} (1093 x-378)
        -\frac{(10159 x+8999)}{81}  H_0
        +\frac{64 \big(199 x^2+174 x+199\big) H_{-1} H_0}{81 (x+1)}\nonumber\\ &&
        -\frac{64}{9} (x+1) H_{-1}^2 H_0
        +\frac{128 \big(x^2+1\big) H_{-1}^3 H_0}{27 (x+1)}
        -\frac{2 \big(4107 x^2+5327 x+3012\big) H_0^2}{81 (x+1)}\nonumber\\ &&
        +\frac{32 \big(19 x^2+18 x+19\big) H_{-1} H_0^2}{27 (x+1)}
        -\frac{32 \big(x^2+1\big) H_{-1}^2 H_0^2}{9 (x+1)}
        +\frac{1}{27} (1503-3905 x) H_1\nonumber\\ &&
        -\frac{2 \big(903 x^2+1126 x+543\big) H_0^3}{81 (x+1)}
        +\frac{64 \big(x^2+1\big) H_{-1} H_0^3}{27 (x+1)}
        -\frac{\big(51 x^2+70 x+51\big) H_0^4}{27 (x+1)}\nonumber\\ &&
        -\frac{32}{81} (319 x+190) H_0 H_1
        -\frac{40}{27} (31 x+16) H_0^2 H_1
        -\frac{152}{27} (x+1) H_0^3 H_1\nonumber\\ &&
        -\frac{8}{9} (3 x+55) H_1^2
        -\frac{8}{9} (9 x+1) H_0 H_1^2
        +\frac{4}{9} (x+1) H_0^2 H_1^2
        -\frac{64}{27} (x+1) H_0 H_1^3\nonumber\\ &&
        -\frac{64 \big(199 x^2+174 x+199\big) H_{0,-1}}{81 (x+1)}
        +\frac{128}{9} (x+1) H_{-1} H_{0,-1}
        +\frac{4}{27} (311 x+467) H_{0,1}\nonumber\\ &&
        -\frac{128 \big(x^2+1\big) H_{-1}^2 H_{0,-1}}{9 (x+1)}
        -\frac{512 \big(4 x^2+3 x+4\big) H_{-1} H_{0,1}}{27 (x+1)}
        +\frac{16}{27} (19 x+52) H_0 H_{0,1}\nonumber\\ &&
        +\frac{16}{3} (x+1) H_0^2 H_{0,1}
        +\frac{64}{9} (x-1) H_1 H_{0,1}
        -\frac{80}{9} (x+1) H_0 H_1 H_{0,1}\nonumber\\ &&
        +\frac{64}{9} (x+1) H_1^2 H_{0,1}
        +\frac{8}{9} (x+1) H_{0,1}^2
        -\frac{128}{9} (x+1) H_{0,-1,-1}\nonumber\\ &&
        +\frac{256 \big(x^2+1\big) H_{-1} H_{0,-1,-1}}{9 (x+1)}
        +\frac{512 \big(4 x^2+3 x+4\big) H_{0,-1,1}}{27 (x+1)}\nonumber\\ &&
        -\frac{64 \big(19 x^2+18 x+19\big) H_{0,0,-1}}{27 (x+1)}
        +\frac{128 \big(x^2+1\big) H_{-1} H_{0,0,-1}}{9 (x+1)}
        -\frac{64}{9} (x+1) H_0 H_{0,0,1}\nonumber\\ &&
        +\frac{4 \big(321 x^2+58 x+57\big) H_{0,0,1}}{27 (x+1)}
        -\frac{128 \big(x^2+1\big) H_{-1} H_{0,0,1}}{9 (x+1)}
        +\frac{128}{9} (x+1) H_1 H_{0,0,1}\nonumber\\ &&
        +\frac{512 \big(4 x^2+3 x+4\big) H_{0,1,-1}}{27 (x+1)}
        -\frac{32}{9} (13 x+1) H_{0,1,1}
        +\frac{256 \big(x^2+1\big) H_{-1} H_{0,1,1}}{9 (x+1)}\nonumber\\ &&
        +\frac{16}{3} (x+1) H_0 H_{0,1,1}
        -\frac{32}{3} (x+1) H_1 H_{0,1,1}
        -\frac{256 \big(x^2+1\big) H_{0,-1,-1,-1}}{9 (x+1)}\nonumber\\ &&
        -\frac{256 \big(x^2+1\big) H_{0,-1,1,1}}{9 (x+1)}
        -\frac{128 \big(x^2+1\big) H_{0,0,-1,-1}}{9 (x+1)}
        +\frac{128 \big(x^2+1\big) H_{0,0,-1,1}}{9 (x+1)}\nonumber\\ &&
        -\frac{128 \big(x^2+1\big) H_{0,0,0,-1}}{9 (x+1)}
        +\frac{8 \big(21 x^2+10 x+21\big) H_{0,0,0,1}}{9 (x+1)}
        +\frac{128 \big(x^2+1\big) H_{0,0,1,-1}}{9 (x+1)}\nonumber\\ &&
        -\frac{32 \big(7 x^2+6 x+7\big) H_{0,0,1,1}}{9 (x+1)}
        -\frac{256 \big(x^2+1\big) H_{0,1,-1,1}}{9 (x+1)}
        -\frac{256 \big(x^2+1\big) H_{0,1,1,-1}}{9 (x+1)}
\bigg]\nonumber\\ &&
+\textcolor{blue}{C_A C_F T_F} \bigg[
        \frac{88}{27} L_M^3 (x+1)
        +\frac{176}{27} L_Q^3 (x+1)
        +L_M^2 \bigg(
                \frac{32}{3} (x+1) H_0
                -\frac{4(83 x-37)}{9} \nonumber\\ &&
                -\frac{32 \big(x^2+1\big) H_{-1} H_0}{3 (x+1)}
                -\frac{16 x H_0^2}{3 (x+1)}
                +\frac{32 \big(x^2+1\big) H_{0,-1}}{3 (x+1)}
                +\frac{32 x \zeta_2}{3 (x+1)}
        \bigg)\nonumber\\ &&
        -L_Q^2 \bigg(
                \frac{4}{27} (997 x+241)
                +\frac{256}{9} (x+1) H_0
                +\frac{32 \big(x^2+1\big) H_{-1} H_0}{3 (x+1)}
                +\frac{16 x H_0^2}{3 (x+1)}\nonumber\\ &&
                +\frac{176}{9} (x+1) H_1
                -\frac{32 \big(x^2+1\big) H_{0,-1}}{3 (x+1)}
                -\frac{32 x \zeta_2}{3 (x+1)}
        \bigg)
        +L_M \bigg(
                -\frac{4}{81} (4577 x-4267)\nonumber\\ &&
                +\bigg(
                        \frac{16 \big(3 x^2+14 x-9\big)}{9 (x+1)}
                        -\frac{64 \big(x^2+1\big) H_{-1}}{3 (x+1)}
                        +\frac{16 \big(3 x^2+4 x+3\big) H_0}{3 (x+1)}
                \bigg) \zeta_2\nonumber\\ &&
                -\frac{16}{27} (29 x-109) H_0
                -\frac{128 \big(4 x^2+3 x+4\big) H_{-1} H_0}{9 (x+1)}
                +\frac{4 \big(19 x^2+4 x+25\big) H_0^2}{9 (x+1)}\nonumber\\ &&
                -\frac{16 \big(x^2+1\big) H_{-1} H_0^2}{3 (x+1)}
                -\frac{32 x H_0^3}{9 (x+1)}
                +\frac{32}{3} (x-1) H_1
                +8 (x+1) H_0^2 H_1\nonumber\\ &&
                +\frac{128 \big(4 x^2+3 x+4\big) H_{0,-1}}{9 (x+1)}
                -\frac{16}{3} (x+1) H_{0,1}
                +\frac{64 \big(x^2+1\big) H_{-1} H_{0,1}}{3 (x+1)}\nonumber\\ &&
                -16 (x+1) H_0 H_{0,1}
                -\frac{64 \big(x^2+1\big) H_{0,-1,1}}{3 (x+1)}
                +\frac{32 \big(x^2+1\big) H_{0,0,-1}}{3 (x+1)}
                +\frac{64 x H_{0,0,1}}{3 (x+1)}\nonumber\\ &&
                -\frac{64 \big(x^2+1\big) H_{0,1,-1}}{3 (x+1)}
                -\frac{32 \big(x^2+3 x+1\big) \zeta_3}{x+1}
        \bigg)
        +L_Q \bigg(
                \frac{4}{81} (19751 x-2371)\nonumber\\ &&
                +\frac{32}{9}\frac{\big(36 x^3+61 x^2+18 x+13\big)H_{-1} H_0}{x+1}  
                -\frac{32}{9}\frac{\big(36 x^3+61 x^2+18 x+13\big)H_{0,-1}}{x+1}\nonumber\\ &&  
                -\frac{8}{9}\frac{\big(72 x^3+39 x^2-89 x-36\big)H_0^2}{x+1}  
                +\bigg(
                         \frac{16}{9}\frac{\big(72 x^3+37 x^2-84 x-29\big)}{x+1} \nonumber\\ &&
                        -\frac{32 \big(7 x^2+6 x+3\big) H_{-1}}{3 (x+1)}
                        -\frac{16 \big(3 x^2+8 x+3\big) H_0}{3 (x+1)}
                        -\frac{64}{3} (4 x+1) H_1
                \bigg) \zeta_2\nonumber\\ &&
                +\frac{32 \big(401 x^2+556 x+137\big) H_0}{27 (x+1)}
                +\frac{80 x H_0^3}{9 (x+1)}
                -\frac{32 \big(7 x^2+6 x+3\big) H_{-1}^2 H_0}{3 (x+1)}\nonumber\\ &&
                +\frac{16 \big(11 x^2+6 x+7\big) H_{-1} H_0^2}{3 (x+1)}
                +\frac{8}{27} (1195 x+169) H_1
                +\frac{352}{9} (x+1) H_0 H_1\nonumber\\ &&
                +\frac{8}{3} (11 x-1) H_0^2 H_1
                +\frac{176}{9} (x+1) H_1^2
                -\frac{16}{3} (x+1) H_0 H_1^2
                +\frac{208}{9} (x+1) H_{0,1}\nonumber\\ &&
                +\frac{64 \big(7 x^2+6 x+3\big) H_{-1} H_{0,-1}}{3 (x+1)}
                +\frac{64 x (3 x+5) H_0 H_{0,-1}}{3 (x+1)}
                -\frac{16}{3} (11 x-1) H_0 H_{0,1}\nonumber\\ &&
                +\frac{64}{3} (x+1) H_1 H_{0,1}
                -\frac{64 \big(7 x^2+6 x+3\big) H_{0,-1,-1}}{3 (x+1)}
                -\frac{32 \big(23 x^2+26 x+7\big) H_{0,0,-1}}{3 (x+1)}\nonumber\\ &&
                +\frac{32}{3} (7 x+1) H_{0,0,1}
                -32 (x+1) H_{0,1,1}
                +\frac{64 \big(3 x^2+5 x+4\big) \zeta_3}{3 (x+1)}
        \bigg)\nonumber\\ &&
        +\bigg(
                \frac{16 \big(174 x^2+209 x-189\big)}{81 (x+1)}
                -\frac{32 \big(29 x^2+18 x+29\big) H_{-1}}{27 (x+1)}
                -\frac{32 \big(x^2+1\big) H_{-1}^2}{9 (x+1)}\nonumber\\ &&
                +\frac{8 \big(63 x^2+29 x+6\big) H_0}{27 (x+1)}
                -\frac{32 \big(x^2+1\big) H_{-1} H_0}{9 (x+1)}
                +\frac{8 \big(3 x^2+8 x+9\big) H_0^2}{9 (x+1)}\nonumber\\ &&
                -\frac{8}{9} (3 x+14) H_1
                +\frac{80}{9} (x+1) H_0 H_1
                +\frac{64}{9} (x+1) H_1^2
                -\frac{16}{9} (7 x+1) H_{0,1}
        \bigg) \zeta_2\nonumber\\ &&
        -\frac{16}{15(x+1)} \big(36 x^2+51 x+22\big) \zeta_2^2
        +\bigg(
                \frac{2 \big(497 x^2+1102 x+1085\big)}{27 (x+1)}
                +\frac{16}{9} (x+1) H_1\nonumber\\ &&
                +\frac{128 \big(x^2+1\big) H_{-1}}{9 (x+1)}
                +\frac{32 \big(6 x^2+4 x-3\big) H_0}{9 (x+1)}
        \bigg) \zeta_3
        +\frac{16}{3} B_4 (x+1)\nonumber\\ &&
        -\frac{4}{81} (995 x-2807) H_0
        -\frac{32 \big(199 x^2+174 x+199\big) H_{-1} H_0}{81 (x+1)}
        +\frac{32}{9} (x+1) H_{-1}^2 H_0\nonumber\\ &&
        -\frac{64 \big(x^2+1\big) H_{-1}^3 H_0}{27 (x+1)}
        +\frac{4 \big(253 x^2+391 x+586\big) H_0^2}{81 (x+1)}
        +\frac{16 \big(x^2+1\big) H_{-1}^2 H_0^2}{9 (x+1)}\nonumber\\ &&
        -\frac{16 \big(19 x^2+18 x+19\big) H_{-1} H_0^2}{27 (x+1)}
        +\frac{8 \big(22 x^2+7 x+25\big) H_0^3}{81 (x+1)}
        -\frac{32 \big(x^2+1\big) H_{-1} H_0^3}{27 (x+1)}\nonumber\\ &&
        -\frac{16 x H_0^4}{27 (x+1)}
        -\frac{8}{27} (65 x-29) H_1
        +\frac{8}{9} (9 x+4) H_0 H_1
        +\frac{8}{9} (14 x+3) H_0^2 H_1\nonumber\\ &&
        +\frac{56}{27} (x+1) H_0^3 H_1
        -\frac{4}{9} (43 x-46) H_1^2
        -\frac{8}{9} (2 x+5) H_0 H_1^2
        -\frac{4}{9} (x+1) H_0^2 H_1^2\nonumber\\ &&
        +\frac{32}{27} (x+1) H_0 H_1^3
        +\frac{32 \big(199 x^2+174 x+199\big) H_{0,-1}}{81 (x+1)}
        -\frac{64}{9} (x+1) H_{-1} H_{0,-1}\nonumber\\ &&
        +\frac{64 \big(x^2+1\big) H_{-1}^2 H_{0,-1}}{9 (x+1)}
        -\frac{8}{27} (143 x+2) H_{0,1}
        +\frac{256 \big(4 x^2+3 x+4\big) H_{-1} H_{0,1}}{27 (x+1)}\nonumber\\ &&
        -\frac{16}{9} (13 x+6) H_0 H_{0,1}
        -\frac{8}{3} (x+1) H_0^2 H_{0,1}
        +\frac{8}{9} (11 x+20) H_1 H_{0,1}\nonumber\\ &&
        +\frac{64}{9} (x+1) H_0 H_1 H_{0,1}
        -\frac{64}{9} (x+1) H_1^2 H_{0,1}
        +\frac{16}{9} (7 x+1) H_{0,1}^2\nonumber\\ &&
        +\frac{64}{9} (x+1) H_{0,-1,-1}
        -\frac{128 \big(x^2+1\big) H_{-1} H_{0,-1,-1}}{9 (x+1)}
        -\frac{256 \big(4 x^2+3 x+4\big) H_{0,-1,1}}{27 (x+1)}\nonumber\\ &&
        +\frac{32 \big(19 x^2+18 x+19\big) H_{0,0,-1}}{27 (x+1)}
        -\frac{64 \big(x^2+1\big) H_{-1} H_{0,0,-1}}{9 (x+1)}\nonumber\\ &&
        +\frac{8 \big(9 x^2+101 x+12\big) H_{0,0,1}}{27 (x+1)}
        +\frac{64 \big(x^2+1\big) H_{-1} H_{0,0,1}}{9 (x+1)}
        -\frac{16}{9} (7 x+1) H_0 H_{0,0,1}\nonumber\\ &&
        -\frac{160}{9} (x+1) H_1 H_{0,0,1}
        -\frac{256 \big(4 x^2+3 x+4\big) H_{0,1,-1}}{27 (x+1)}
        -\frac{16}{9} (x+7) H_{0,1,1}\nonumber\\ &&
        -\frac{128 \big(x^2+1\big) H_{-1} H_{0,1,1}}{9 (x+1)}
        -\frac{16}{9} (11 x+5) H_0 H_{0,1,1}
        +16 (x+1) H_1 H_{0,1,1}\nonumber\\ &&
        +\frac{128 \big(x^2+1\big) H_{0,-1,-1,-1}}{9 (x+1)}
        +\frac{128 \big(x^2+1\big) H_{0,-1,1,1}}{9 (x+1)}
        +\frac{64 \big(x^2+1\big) H_{0,0,-1,-1}}{9 (x+1)}\nonumber\\ &&
        -\frac{64 \big(x^2+1\big) H_{0,0,-1,1}}{9 (x+1)}
        +\frac{64 \big(x^2+1\big) H_{0,0,0,-1}}{9 (x+1)}
        +\frac{64 \big(5 x^2+6 x-1\big) H_{0,0,0,1}}{9 (x+1)}\nonumber\\ &&
        -\frac{64 \big(x^2+1\big) H_{0,0,1,-1}}{9 (x+1)}
        -\frac{16 \big(x^2-2 x-11\big) H_{0,0,1,1}}{9 (x+1)}
        +\frac{128 \big(x^2+1\big) H_{0,1,-1,1}}{9 (x+1)}\nonumber\\ &&
        +\frac{128 \big(x^2+1\big) H_{0,1,1,-1}}{9 (x+1)}
        -\frac{112}{9} (x+1) H_{0,1,1,1}
        -\frac{2}{729} (108295 x-86681)
\bigg]\nonumber\\ &&
+\textcolor{blue}{C_F T_F^2} \bigg[
        -\frac{64}{27} L_M^3 (x+1)
        -\frac{32}{27} L_Q^3 (x+1)
        -L_M^2 \bigg(
                \frac{32}{27} (11 x-1)\nonumber\\ &&
                +\frac{32}{9} (x+1) H_0
        \bigg)
        +L_Q^2 \bigg(
                \frac{32}{27} (17 x+8)
                +\frac{64}{9} (x+1) H_0
                +\frac{32}{9} (x+1) H_1
        \bigg)\nonumber\\ &&
        -\frac{992}{81} L_M (x+1)
        +L_Q \bigg(
                -\frac{32}{81} (280 x+37)
                -\frac{64}{27} (34 x+19) H_0
                -\frac{32}{3} (x+1) H_0^2\nonumber\\ &&
                -\frac{64}{27} (17 x+8) H_1
                -\frac{64}{9} (x+1) H_0 H_1
                -\frac{32}{9} (x+1) H_1^2
                -\frac{64}{9} (x+1) H_{0,1}\nonumber\\ &&
                +\frac{128}{9} (x+1) \zeta_2
        \bigg)
        +\frac{64}{81} (6 x-7) H_0
        +\frac{16}{81} (11 x-1) H_0^2
        +\frac{16}{81} (x+1) H_0^3\nonumber\\ &&
        -\frac{448}{27} (x+1) \zeta_3
        +\frac{16}{729} (431 x+323)
\bigg]\nonumber\\ &&
+\textcolor{blue}{C_F N_F T_F^2} \bigg[
        -\frac{32}{27} L_M^3 (x+1)
        -\frac{64}{27} L_Q^3 (x+1)
        +L_Q^2 \bigg(
                \frac{64}{27} (17 x+8)\nonumber\\ &&
                +\frac{128}{9} (x+1) H_0
                +\frac{64}{9} (x+1) H_1
        \bigg)
        +L_M \bigg(
                \frac{32}{81} (5 x-73)
                +\frac{32}{27} (11 x-1) H_0\nonumber\\ &&
                +\frac{16}{9} (x+1) H_0^2
        \bigg)
        +L_Q \bigg(
                -\frac{64}{81} (280 x+37)
                -\frac{128}{27} (34 x+19) H_0
                -\frac{64}{3} (x+1) H_0^2\nonumber\\ &&
                -\frac{128}{27} (17 x+8) H_1
                -\frac{128}{9} (x+1) H_0 H_1
                -\frac{64}{9} (x+1) H_1^2
                -\frac{128}{9} (x+1) H_{0,1}\nonumber\\ &&
                +\frac{256}{9} (x+1) \zeta_2
        \bigg)
        +\frac{128}{81} (6 x-7) H_0
        +\frac{32}{81} (11 x-1) H_0^2
        +\frac{32}{81} (x+1) H_0^3\nonumber\\ &&
        +\frac{256}{27} (x+1) \zeta_3
        -\frac{64}{729} (161 x+215)
\bigg]
\bigg\}+\hat{c}^{(3)}_{q,2}~.
\end{eqnarray}

Here the $+$-prescription is defined by
\begin{equation}
\int_0^1 dx g(x) [f(x)]_+ = \int_0^1 dx [g(x)-g(1)]f(x)~.
\end{equation}

The contribution of the massive Wilson coefficient $H_{q,2}$ 
is found by combining the massless Wilson coefficient $C_{q,2}$ and $L_{q,2}$ by
\begin{equation}
H_{q,2}(x,Q^2) = C_{q,2}(N_F,x,Q^2)+L_{q,2}(x,Q^2)~.
\end{equation}
Except for $c^{(3)}_{q,2}(N_F)$, the Wilson coefficients are expressed by up to weight {\sf w=4} harmonic 
polylogarithms. Note the emergence of a denominator $1/(1-x)^2$, cf.~\cite{Ablinger:2014vwa}, which is 
properly regularized by its numerator function in the limit $x \rightarrow 1$.  We note that we have applied 
the shuffle algebra, cf. 
\cite{Blumlein:2003gb}, which leads to a reduction of the number of harmonic polylogarithms compared to the
linear representation, making the numerical evaluation faster.
\section{Numerical Results}
\label{sec:5}

\vspace*{1mm}
\noindent
In the following we illustrate the asymptotic charm corrections up to 3-loop order to the
charged current non-singlet combinations $F_{1,2}^{W^+-W^-}(x,Q^2)$ choosing the renormalization and 
factorization scales $\mu^2  = Q^2$. First we consider the behaviour of the corrections at small and large 
values of the Bjorken variable $x$. For those of the massless 3-loop Wilson coefficients 
see \cite{Davies:2016ruz}.
The limiting behaviour for the two contributing functions 
$L_{q,i}^{W^+-W^-,\rm NS}(N_F+1) - \hat{C}_{q,i}^{W^+-W^-,\rm NS}(N_F)$ and
$H_{q,i}^{W^+-W^-,\rm NS}(N_F+1) - {C}_{q,i}^{W^+-W^-,\rm NS}(N_F+1)$ are the same, see also 
\cite{Blumlein:2016xcy}.

For the 3-loop contributions, yet for general values of $\mu^2$, at low values of $x$ one has
\begin{eqnarray}
L_{q,L}^{W^+-W^-\rm NS}(N_F+1)-\hat{C}_{q,L}^{W^+-W^-,\rm NS}(N_F) &\propto& a_s^3\,\frac{8}{3}C_F^2 T_F\, 
\ln^2(x)
\\
L_{q,2}^{W^+-W^-,\rm NS}(N_F+1)-\hat{C}_{q,2}^{W^+-W^-,\rm NS}(N_F) &\propto& 
a_s^3 \Biggl\{
\frac{16}{27} C_A C_F T_F - \frac{5}{9} C_F^2 T_F 
\Biggr\} \ln^4(x)
\end{eqnarray}
and at large $x$
\begin{eqnarray}
L_{q,L}^{W^+-W^-,\rm NS}(N_F+1)-\hat{C}_{q,L}^{W^+-W^-,\rm NS}(N_F)
&\propto&
a_s^3 C_F T_F \Biggl\{\frac{128}{3} C_F L_Q \ln^2(1-x)
+ \Biggl[C_F \Biggl(\frac{896}{27}  
\nonumber\\ &&
+ \frac{320}{9}  L_M 
+ \frac{32}{3} L_M^2 \Biggr) 
+ 32 C_F L_Q^2 + 
    \Biggl(- \frac{1088}{9} C_A 
\nonumber\\ &&
+ \frac{80}{9} C_F + \frac{128}{9} T_F 
+ \frac{256}{9} N_F T_F + 
\frac{128}{3} \zeta_2 C_A 
\nonumber\\ &&
- \frac{256}{3} \zeta_2 C_F \Biggr) L_Q \Biggr] \ln(1-x) 
\Bigg\}
\\
L_{q,2}^{W^+-W^-,\rm NS}(N_F+1)-\hat{C}_{q,2}^{W^+-W^-,\rm NS}(N_F)
&\propto& a_s^3 C_F T_F \Biggl\{
\frac{320}{9} C_F  L_Q  \left(\frac{\ln^3(1-x)}{1-x}\right)_+
\nonumber\\ &&
+  \Biggl[C_F \Biggl(\frac{448}{9}  + \frac{160}{3} L_M 
+ 16  L_M^2\Biggr) + 48 C_F L_Q^2 
\nonumber\\ &&
+ 
   \Biggl(-\frac{352}{9} C_A - \frac{512}{3} C_F + \frac{64}{9} T_F 
\nonumber\\ &&
+ \frac{128}{9} N_F T_F\Biggr) 
L_Q \Biggr]  \left(\frac{\ln^2(1-x)}{1-x}\right)_+\Biggr\}~.
\end{eqnarray}

Below we plot the heavy flavor contribution to the structure function $F_1^{W^+-W^-}(x,Q^2)$ 
for the quark mass $m_c=1.59\, \GeV$ in the on-shell scheme \cite{Alekhin:2012vu} and the scales 
$Q^2=\mu^2=10,\,100,\,1000\,~\GeV^2$ for the complete structure function, including the massive and 
massless terms.

\vspace*{-5mm}
\begin{figure}[H]
\begin{center}
\includegraphics[width=0.7\linewidth]{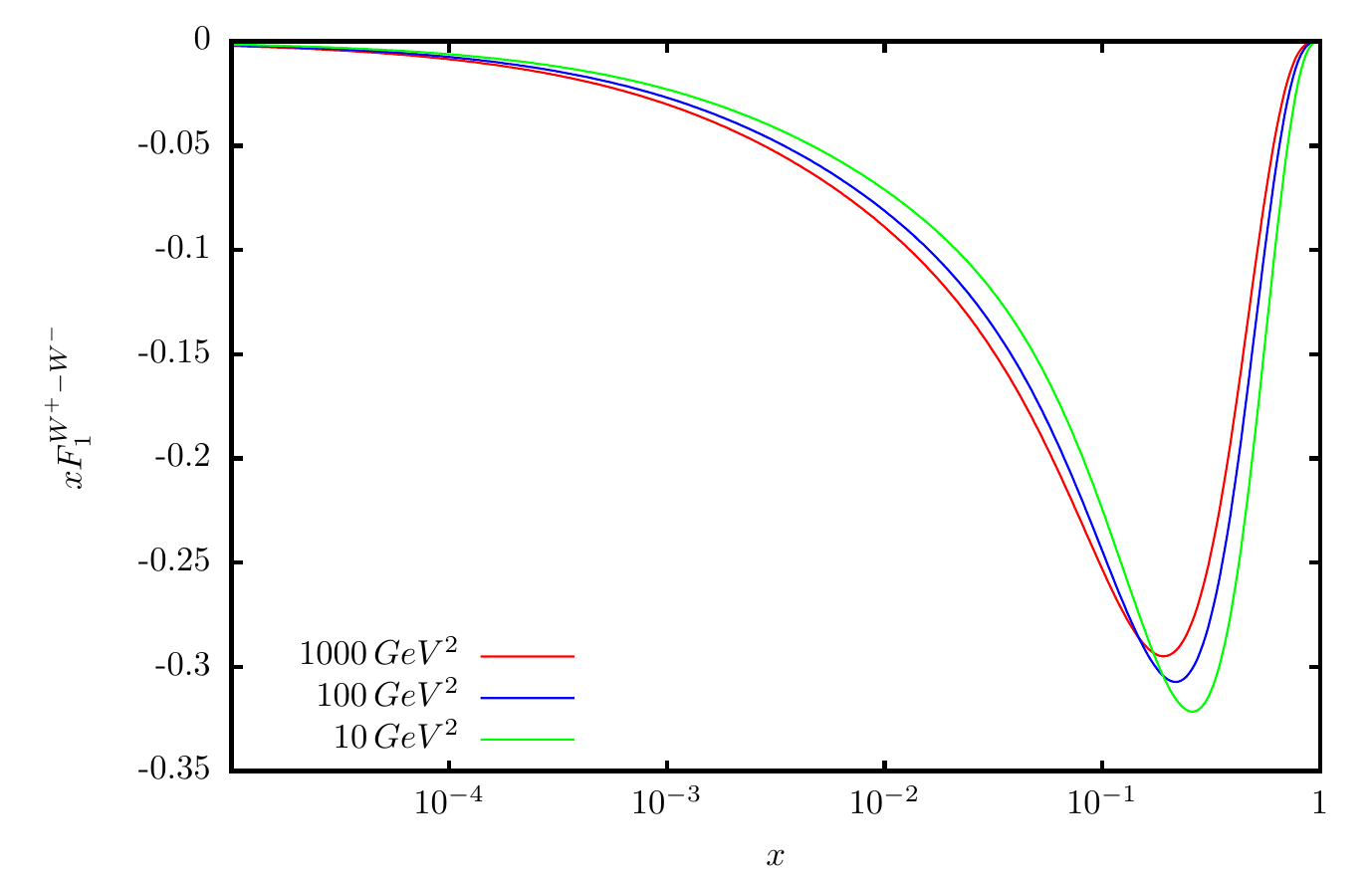}
\end{center}
\caption{\sf The structure function $xF_1^{W^+-W^-}(x,Q^2)$, containing the 3-loop corrections including the 
asymptotic corrections for charm using $m_c^{\rm OMS} = 1.59~\GeV$ and the PDFs \cite{Alekhin:2013nda}.}
\label{fig:1}
\end{figure}
\begin{figure}[H]
\begin{center}
\includegraphics[width=0.7\linewidth]{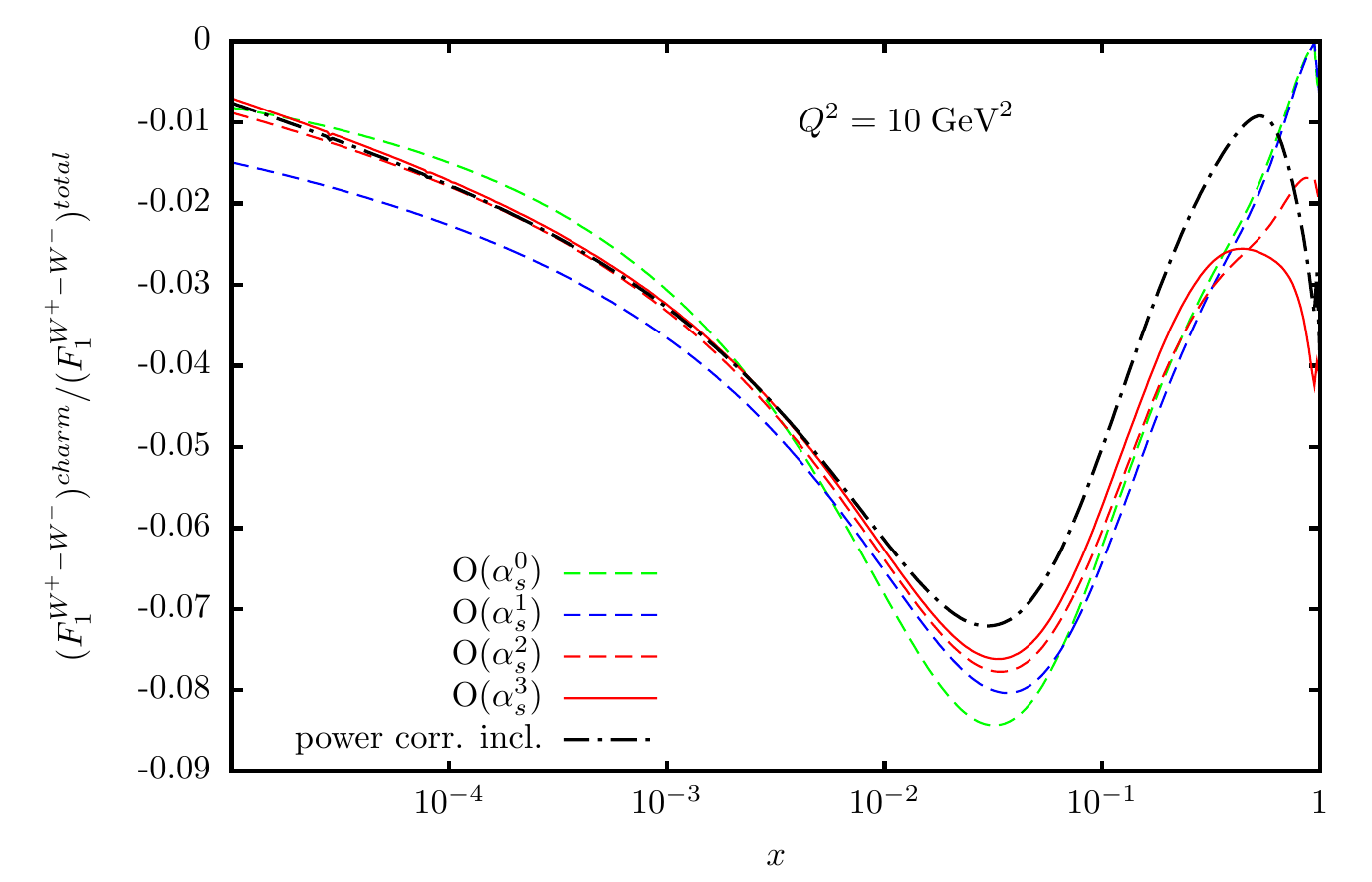}
\end{center}
\caption{\sf The ratio of massive contributions to the structure function $xF_1^{W^+-W^-}(x,Q^2)$ over the complete 
structure function for $Q^2=10~\GeV^2$, containing the 3-loop corrections including the asymptotic corrections for 
charm using $m_c^{\rm OMS} = 1.59~\GeV$ and the PDFs \cite{Alekhin:2013nda}. 
For the dash-dotted line, asymptotic corrections at three loops and the complete
heavy flavor contributions up to $O(a_s^2)$ \cite{Blumlein:2016xcy}
are taken into account.}
\label{fig:2}
\end{figure}

In Figure~\ref{fig:1} the scale evolution of the structure function $xF_1^{W^+-W^-}(x,Q^2)$ is shown in the range
$Q^2 \in [10,1000]~\GeV^2$, including the asymptotic charm quark corrections to 3-loop order. 

\vspace*{-3mm}
\begin{figure}[H]
\begin{center}
\includegraphics[width=0.7\linewidth]{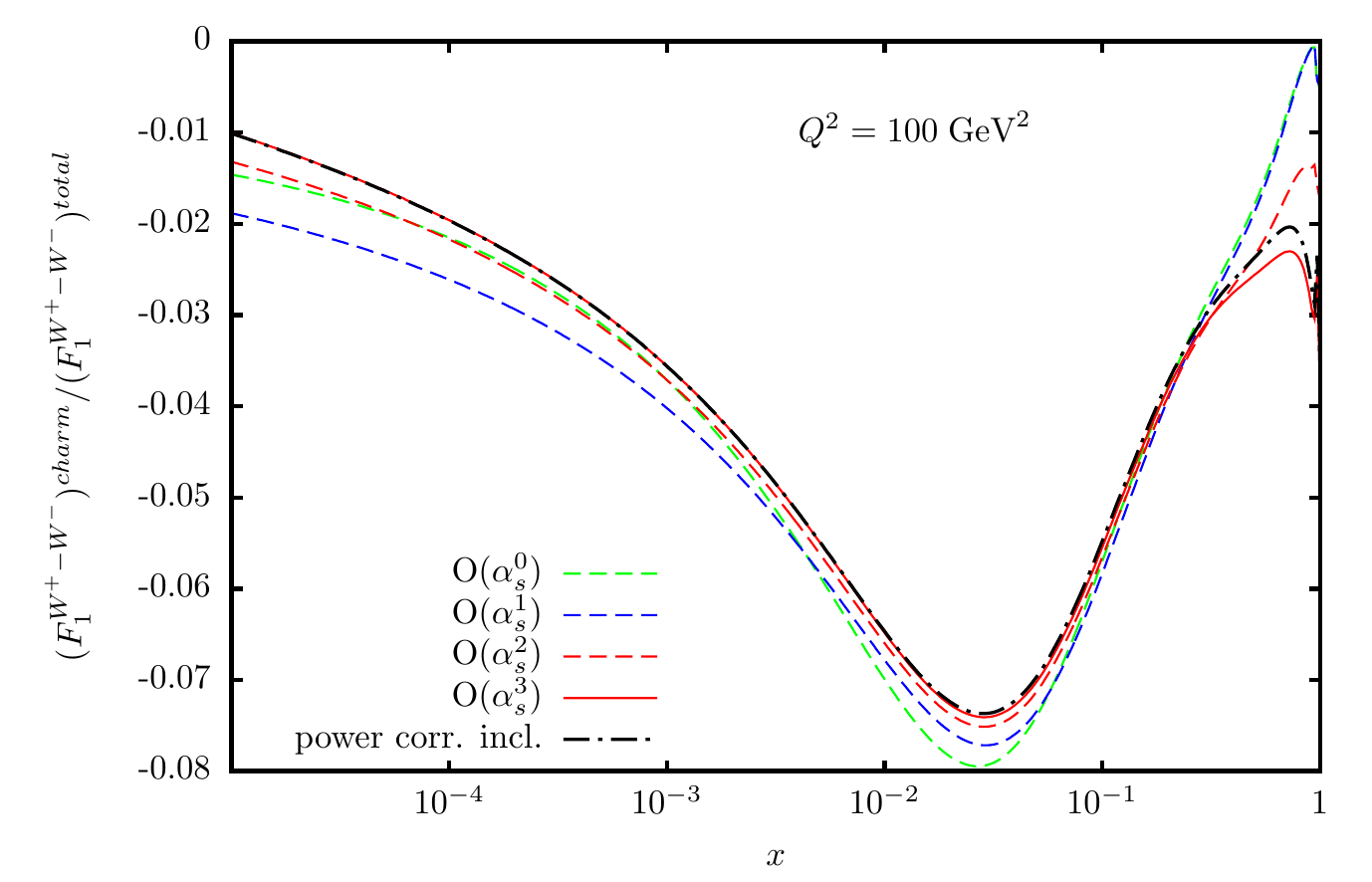}
\end{center}
\caption{\sf The ratio of massive contributions to the structure function $xF_1^{W^+-W^-}(x,Q^2)$ over the complete 
structure function for $Q^2=100~\GeV^2$, containing the 3-loop corrections including the asymptotic corrections for 
charm using $m_c^{\rm OMS} = 1.59~\GeV$ and the PDFs \cite{Alekhin:2013nda}. 
For the dash-dotted line, asymptotic corrections at three loops and the complete
heavy flavor contributions up to $O(a_s^2)$ \cite{Blumlein:2016xcy}
are taken into account.}
\label{fig:3}
\end{figure}

\vspace*{-3mm}
\begin{figure}[H]
\begin{center}
\includegraphics[width=0.7\linewidth]{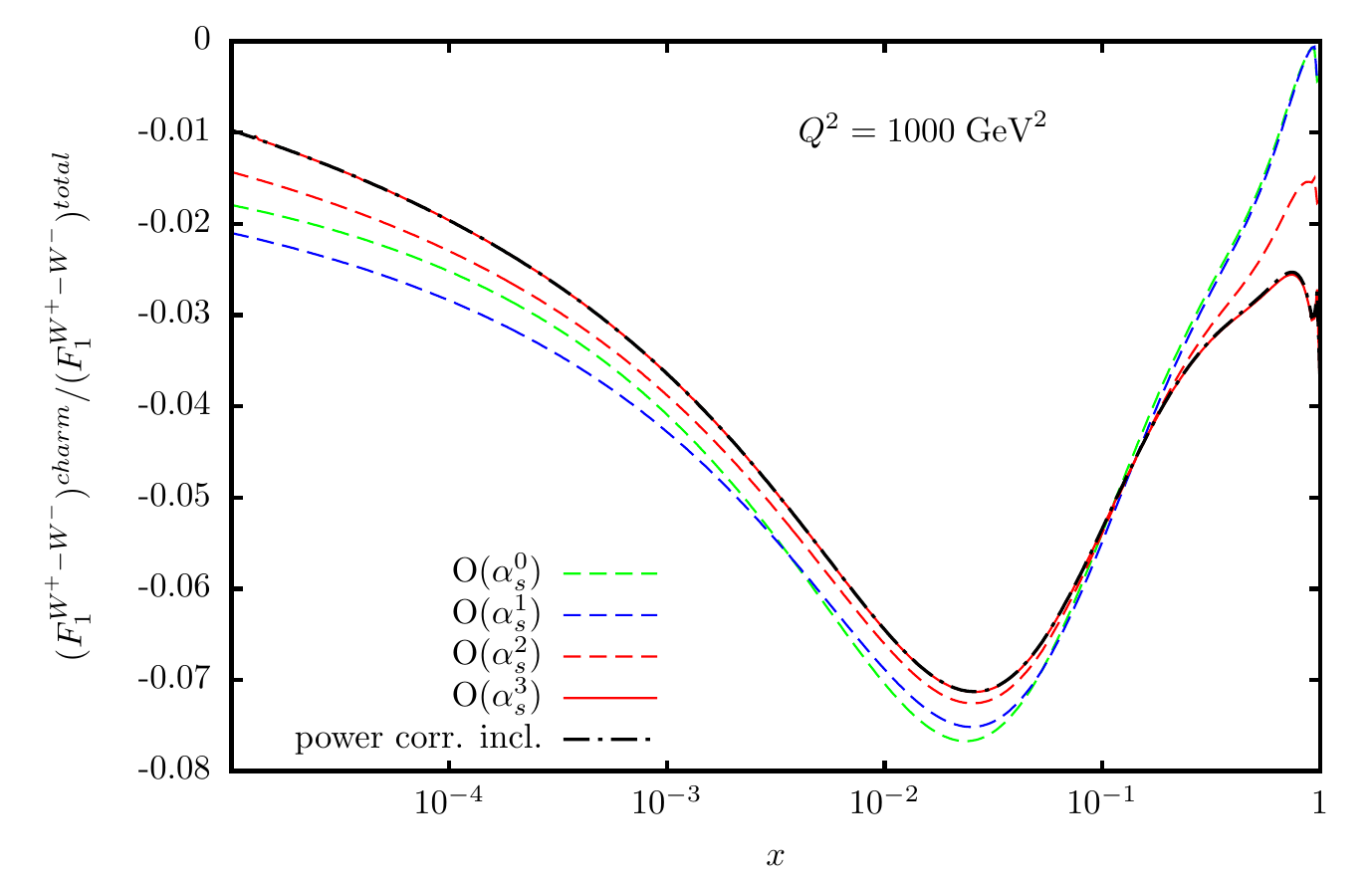}
\end{center}
\caption{\sf The ratio of massive contributions to the structure function $xF_1^{W^+-W^-}(x,Q^2)$ over the complete 
structure function for $Q^2=1000~\GeV^2$, containing the 3-loop corrections including the asymptotic corrections for 
charm using $m_c^{\rm OMS} = 1.59~\GeV$ and the PDFs \cite{Alekhin:2013nda}.
For the dash-dotted line, asymptotic corrections at  three loops and the complete
heavy flavor contributions up to $O(a_s^2)$ \cite{Blumlein:2016xcy}
are taken into account.}
\label{fig:4}
\end{figure}

\noindent
Here and in the 
following we refer to the parton distribution functions~\cite{Alekhin:2013nda}. As typical for non-singlet 
contributions, the profile is shifted from larger to smaller values of $x$ with growing values of $Q^2$. However,
the effects are much smaller than in the  singlet case. As it is well known, the validity of the asymptotic 
charm quark corrections in the case of $F_L(x,Q^2)$, and therefore for $F_2$ and in part for 
$xF_1$, is setting in at higher scales only due to the $F_L$ contribution, for details see \cite{Buza:1995ie}.
We will discuss these aspects in the following figures for $xF_1$ and $F_2$.

In Figure~\ref{fig:2} the corrections to $xF_1^{W^+-W^-}(x,Q^2)$ are illustrated for $Q^2 = 10~\GeV^2$ by 
adding the contributions from $O(a_s^0)$ to  $O(a_s^3)$, showing  an increasing degree of 
stabilization.  We also present
the exact heavy flavor corrections to $O(a_s^2)$ \cite{Blumlein:2016xcy}, showing deviations in the range
$x \gsim 10^{-2}$, while below there is exact agreement. The latter effect is due to the sufficiently large $W^2 = 
Q^2 (1-x)/x$ values through which the heavy quarks are made effectively massless for this structure function
even at this low scale of $Q^2$. The charm quark corrections for $xF_1^{W^+-W^-}(x,Q^2)$ vary in a range of 
$-8 \%$ to $\sim 0\%$, depending on $x$, with a maximal relative contribution around $x \sim 3 \cdot 
10^{-2}$.

Figure~\ref{fig:3} shows that at $Q^2 = 100~\GeV^2$ the asymptotic corrections agree also in the case where we 
include the power corrections to larger values of $x \sim 0.3$ and for $Q^2 = 1000~\GeV^2$, Figure~\ref{fig:4},
the agreement is obtained in the whole $x$ range.

\begin{figure}[H]
\begin{center}
\includegraphics[width=0.7\linewidth]{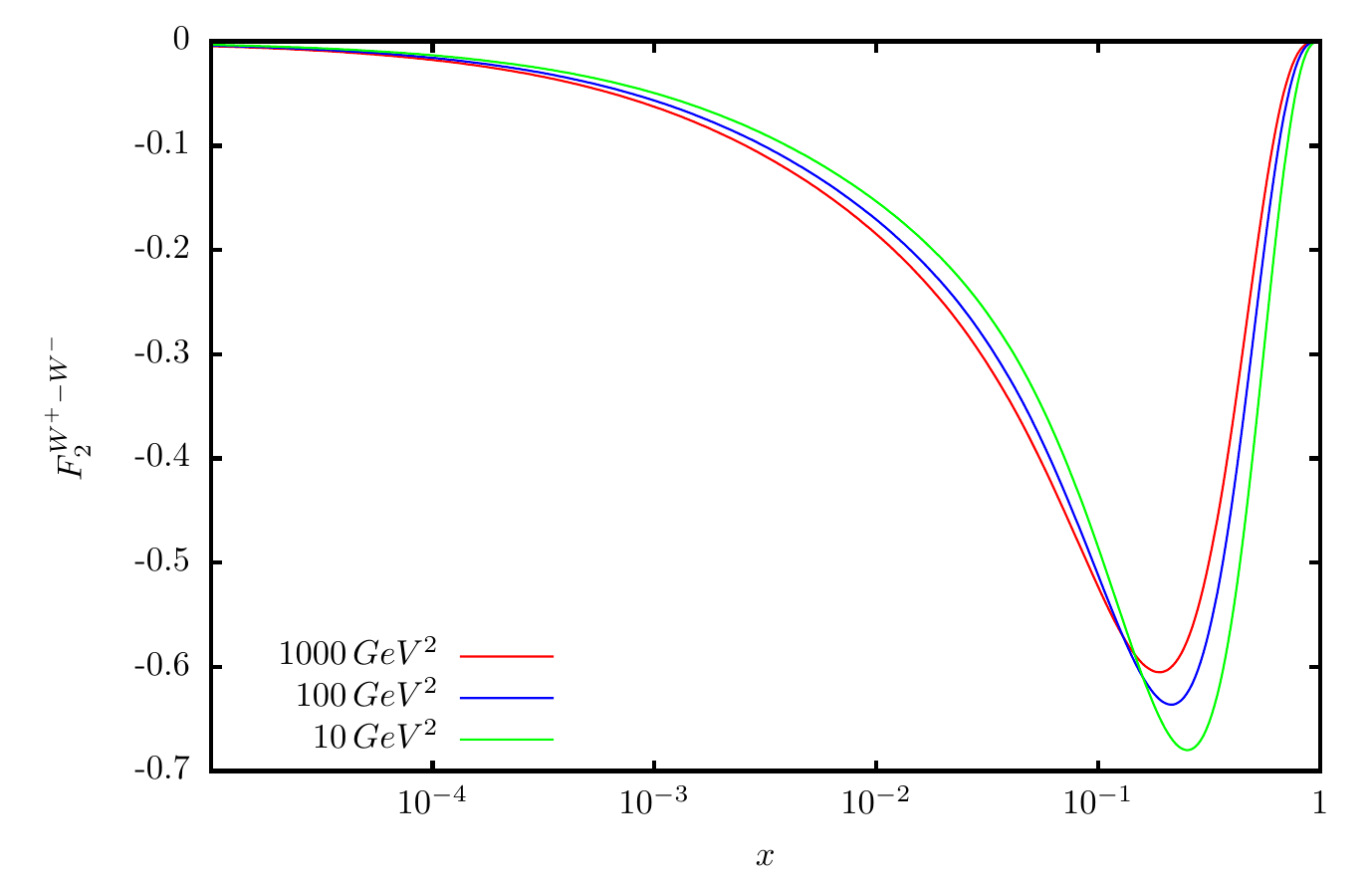}
\end{center}
\caption{\sf The structure function $F_2^{W^+-W^-}(x,Q^2)$, containing the 3-loop corrections including the
asymptotic corrections for charm using $m_c^{\rm OMS} = 1.59~\GeV$ and the PDFs \cite{Alekhin:2013nda}.}
\label{fig:5}
\end{figure}

\begin{figure}[H]
\begin{center}
\includegraphics[width=0.7\linewidth]{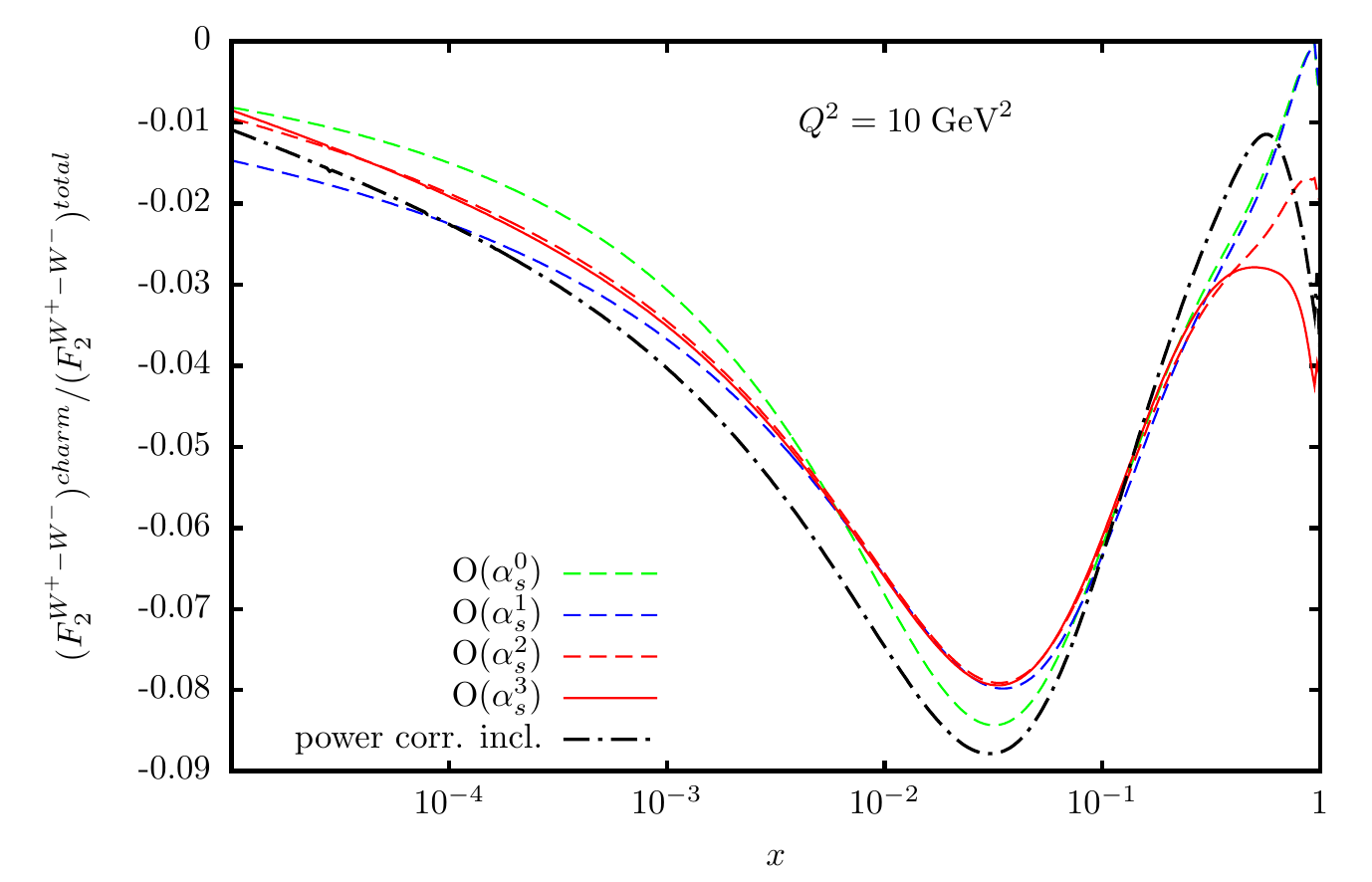}
\end{center}
\caption{\sf The ratio of massive contributions to the structure function $F_2^{W^+-W^-}(x,Q^2)$ over the complete
structure function for $Q^2=10~\GeV^2$, containing the 3-loop corrections including the asymptotic corrections for
charm using $m_c^{\rm OMS} = 1.59~\GeV$ and the PDFs \cite{Alekhin:2013nda}.
For the dash-dotted line, asymptotic corrections at three loops and the complete
heavy flavor contributions up to $O(a_s^2)$ \cite{Blumlein:2016xcy}
are taken into account.}
\label{fig:6}
\end{figure}

We turn now to the numerical illustration of the structure function $F_2^{W^+-W^-}(x,Q^2)$.
In Figure~\ref{fig:5} we show the scaling violations of $F_2^{W^+-W^-}(x,Q^2)$ in the region $Q^2  \in 
[10,1000]~\GeV^2$, shifting the profile to lower values of $x$ with growing virtualities $Q^2$. Figure~\ref{fig:6}
shows the contributions to $F_2^{W^+-W^-}(x,Q^2)$ at $Q^2 = 10~\GeV^2$ for growing order in the strong coupling 
constant stabilizing at 3-loop order, except of very large values of $x$. At $Q^2 = 10~\GeV^2$ comparing the 
results for $2xF_1$ and $F_2$ the effect of $F_L(x,Q^2)$ is clearly visible. The asymptotic expression is 
not yet 
valid in the charged current case, as the complete $O(a_s^2)$ charm quark corrections show. 

\begin{figure}[H]
\begin{center}
\includegraphics[width=0.7\linewidth]{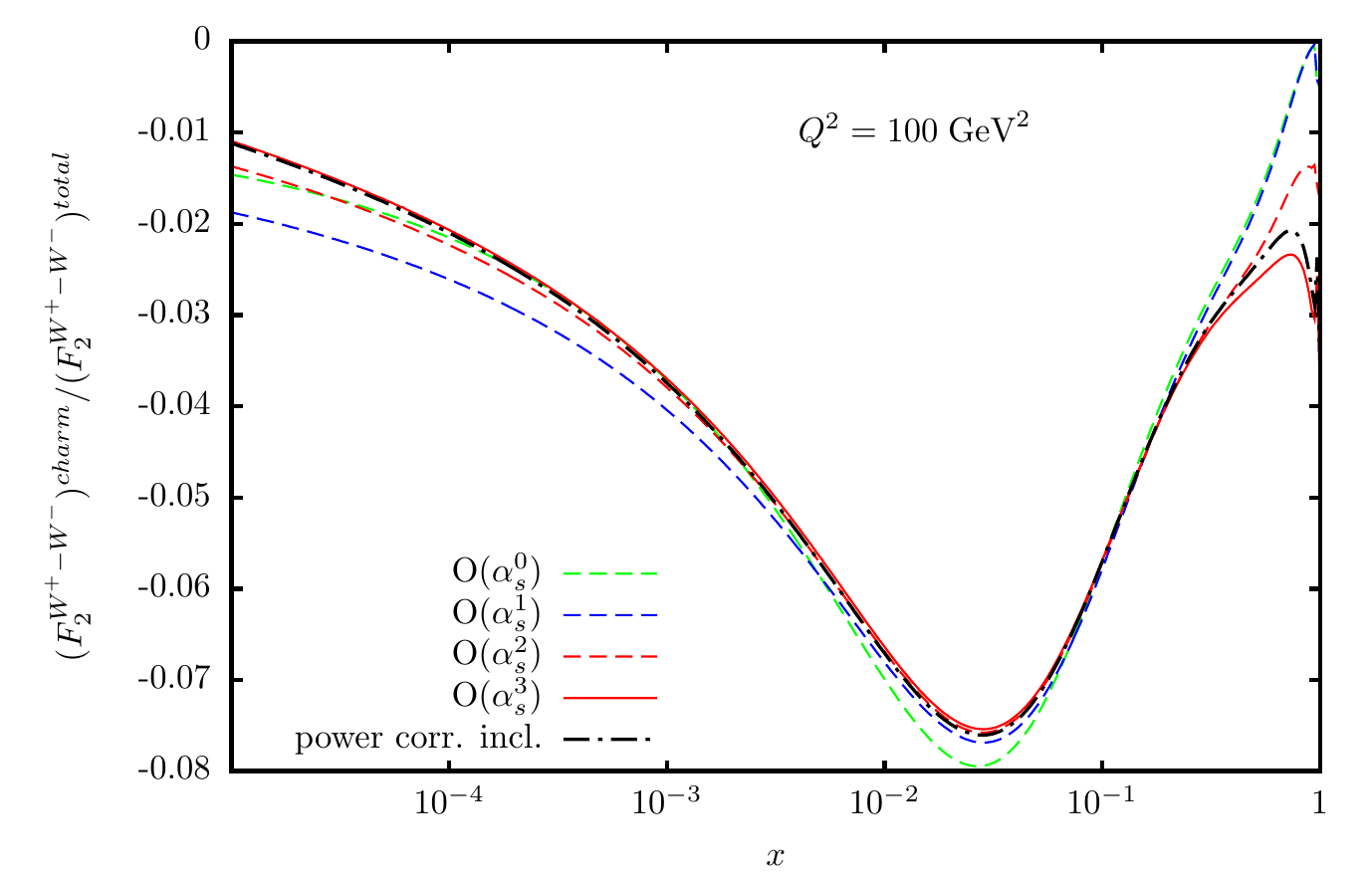}
\end{center}
\caption{\sf The ratio of massive contributions to the structure function $F_2^{W^+-W^-}(x,Q^2)$ over the complete
structure function for $Q^2=100~\GeV^2$, containing the 3-loop corrections including the asymptotic corrections for
charm using $m_c^{\rm OMS} = 1.59~\GeV$ and the PDFs \cite{Alekhin:2013nda}.
For the dash-dotted line, asymptotic corrections at three loops and the complete
heavy flavor contributions up to $O(a_s^2)$ \cite{Blumlein:2016xcy}
are taken into account.}
\label{fig:7}
\end{figure}

\begin{figure}[H]
\begin{center}
\includegraphics[width=0.7\linewidth]{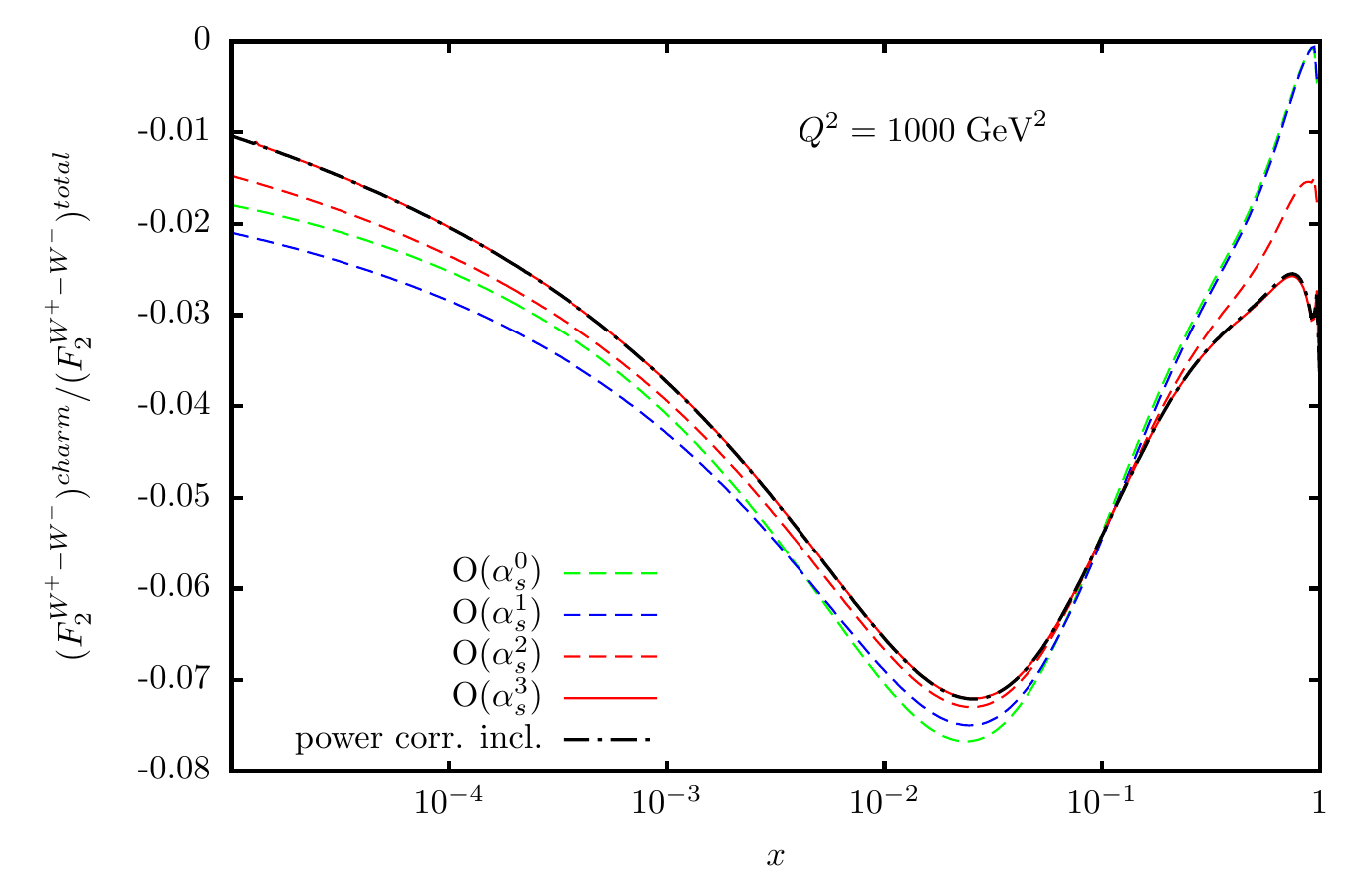}
\end{center}
\caption{\sf The ratio of massive contributions to the structure function $F_2^{W^+-W^-}(x,Q^2)$ over the complete
structure function for $Q^2=1000~\GeV^2$, containing the 3-loop corrections including the asymptotic corrections for
charm using $m_c^{\rm OMS} = 1.59~\GeV$ and the PDFs \cite{Alekhin:2013nda}.
For the dash-dotted line, asymptotic corrections at three loops and the complete
heavy flavor contributions up to $O(a_s^2)$ \cite{Blumlein:2016xcy}
are taken into account.}
\label{fig:8}
\end{figure}

\noindent
Again the relative 
charm quark corrections vary in the range $[-8\%, \sim 0\%]$. As shown in Figure~\ref{fig:7}, the asymptotic 
corrections agree with the case where the power corrections are included, except for a small range at very large 
$x$ values 
at $Q^2 = 100~\GeV^2$. Finally, this effect disappears for $Q^2 = 1000~\GeV^2$, see Figure~\ref{fig:8}.


\section{The Sum Rules}
\label{sec:6}

\vspace*{1mm}
\noindent
For the combination of the charged current structure functions being considered here, there exist sum rules 
arising from the lowest Mellin moment. In the case of $F_2^{W^+-W^-}(x,Q^2)$, one obtains the Adler sum rule 
\cite{Adler:1965ty}
and for $F_1^{W^+-W^-}(x,Q^2)$ the unpolarized Bjorken sum rule \cite{Bjorken:1967px}, for which also the 
target mass corrections have to be considered, cf. \cite{Blumlein:2016xcy}.

The Adler sum rule states
\begin{eqnarray}
\label{eq:ADLER}
\int_0^1 \frac{dx}{x} \left[F_2^{\overline{\nu}p}(x,Q^2) - F_2^{{\nu}p}(x,Q^2)\right] = 2 [1 + \sin^2(\theta_c)]
\end{eqnarray}
for three massless flavors. Here $\theta_c$ denotes the Cabibbo angle \cite{Cabibbo:1963yz}. The integral 
(\ref{eq:ADLER}) neither receives QCD nor quark- or target mass corrections \cite{Blumlein:2012bf}, cf. also 
\cite{Ravindran:2001dk,Adler:2009dw}. Up to 2-loop order the vanishing of the heavy quark corrections has been 
shown 
in Ref.~\cite{Blumlein:2016xcy}. Considering the limit of large scales $Q^2 \gg m^2$, this is confirmed at 3-loop 
order
since the flavor non-singlet OMEs vanish for $N=1$ due to fermion number conservation \cite{Ablinger:2014vwa} and the 
first moment of the corresponding massless Wilson coefficient also vanishes \cite{Moch:2007gx}.

The unpolarized Bjorken sum rule \cite{Bjorken:1967px} is given by
\begin{eqnarray}
\label{eq:UPBJ}
\int_0^1 {dx} \left[F_1^{\bar{\nu} p}(x,Q^2) - F_1^{{\nu} p}(x,Q^2)\right] 
= C_{\rm uBJ}(\hat{a}_s),
\end{eqnarray}
with $\hat{a}_s = \alpha_s/\pi$.
The massless 1-loop 
\cite{Bardeen:1978yd,Altarelli:1978id,Humpert:1980uv,Furmanski:1981cw}, 2-loop \cite{Gorishnii:1985xm}, 3-loop 
\cite{Larin:1990zw} and 4-loop 
\cite{Chetyrkin:14} QCD corrections have been calculated 
\begin{eqnarray}
\label{eq:uBJSR}
C_{\rm uBJ}(\hat{a}_s),
&=& 
1 -  0.66667 \hat{a}_s 
+ \hat{a}_s^2 (-3.83333 + 0.29630 N_F)
\nonumber\\ &&
+ \hat{a}_s^3 (-36.1549 + 6.33125 N_F - 0.15947 N_F^2) 
\nonumber\\ &&
+ \hat{a}_s^4 (-436.768 + 111.873 N_F - 7.11450 N_F^2 + 0.10174 N_F^3)~,
\end{eqnarray} 
setting $\mu^2 = Q^2$ for $SU(3)_c$.
For $N_F = 3, 4$ the massless QCD corrections are given by
\begin{eqnarray}
\label{eq:uBJSRM0}
C_{\rm uBJ}(\hat{a}_s, N_F=3)
&=& 
1 -  0.66667 \hat{a}_s  - 2.94444 \hat{a}_s^2 - 18.5963 \hat{a}_s^3 - 162.436 \hat{a}_s^4 
\\
C_{\rm uBJ}(\hat{a}_s, N_F=4) &=&
1 -  0.66667 \hat{a}_s  - 2.64815 \hat{a}_s^2 - 13.3813 \hat{a}_s^3 - 96.6032 \hat{a}_s^4~. 
\end{eqnarray} 
The massive corrections start at $O(a_s^0)$ with the $s'= (|V_{dc}|^2 d
+ |V_{sc}|^2 s) \rightarrow c$ transitions \cite{Gluck:1997sj,Blumlein:2011zu} and have been given in 
complete form in Ref.~\cite{Blumlein:2016xcy} to 2-loop order. The charm corrections at $O(\hat{a}_s^2)$ 
are of the same size as the massless $O(\hat{a}_s^4)$ corrections. Ref.~\cite{Blumlein:2016xcy} also 
contains the target mass corrections.
In the asymptotic case, the effect of the heavy flavor corrections reduces to a shift of $N_F \rightarrow N_F +1$
in the massless corrections since the massive OMEs vanish for $N=1$ due to fermion conservation, which holds to all
orders in perturbation theory.

\section{Conclusions}
\label{sec:7}

\vspace*{1mm}
\noindent
We have calculated the massive charm  quark 3-loop corrections to  the charged current Wilson 
coefficients for the structure functions $F_{1,2}^{W^+ - W^-}(x,Q^2)$ in the asymptotic region 
$Q^2 \gg m_c^2$ both in Mellin $N$ and $x$ space. The corresponding contributions are composed 
of two massive Wilson coefficients $L_{q}^{W^+-W^-,\rm NS}$ and $H_{q}^{W^+-W^-,\rm NS}$ for 
which the weak boson either 
couples to a massless ($L$) or a 
massive quark line ($H$), here in the $s' \rightarrow c$ transition. The massless 3-loop Wilson 
coefficients have been calculated in \cite{Davies:2016ruz} and the massive OMEs were presented before
in \cite{Ablinger:2014vwa} as part of the present project to compute all massive 3-loop corrections
to deep-inelastic scattering at high values of $Q^2$. The 
results have a representation in terms of 
nested harmonic sums and harmonic polylogarithms only. The charm quark corrections in case of both 
structure functions amount up to $\sim 8\%$, depending on $x$ and the 3-loop corrections stabilize
lower order QCD results. At low values of $Q^2$, effects of power corrections are still visible, which we 
have illustrated using recent complete  2-loop results \cite{Blumlein:2016xcy}, while for $Q^2 \gsim 
100~\GeV^2$ the asymptotic representation is valid in a rather wide range of $x$.

We also discussed potential contributions of the present corrections to the Adler and unpolarized 
Bjorken sum rules. In the former case, in accordance with the expectation, no corrections are obtained.
For the Bjorken sum rule, the charm quark contributions lead to  a shift of $N_F = 3$ by one unit in the 
massless result. There are no heavy quark contributions due to fermion number conservation, which is 
expressed by a vanishing first moment of the operator matrix element in the 
non-singlet cases. Therefore, only the massless terms contribute now with $N_F \rightarrow N_F + 1$.

The 3-loop charm quark corrections to the structure functions $F_{1,2}^{W^+ - W^-}(x,Q^2)$ will improve 
the analysis of the HERA charged current data and are relevant for precision measurements in 
deep-inelastic scattering at planned facilities like the EIC \cite{EIC}, LHeC \cite{LHEC} and neutrino 
factories \cite{NUFACT} in the future, which will reach a higher statistical and systematic precision 
than obtained in present experiments.

\vspace{5mm}
\noindent
{\bf Acknowledgment.}~
We would like to thank J.~Ablinger, A.~Hasselhuhn and A.~Vogt for discussions, as well  A.~Vogt for providing
us the effective numerical representations of the massless flavor non-singlet 3-loop Wilson coefficients of 
Ref.~\cite{Davies:2016ruz}. This work was supported in part by the European Commission through contract 
PITN-GA-2012-316704 ({HIGGSTOOLS}) and the Austrian Science Fund (FWF) grant SFB F50 (F5009-N15).

\newpage

\end{document}